\documentclass[symmetry,article,accept,oneauthor,pdftex]{mdpi} 
\pdfoutput=1
\firstpage{1} 
\pubvolume{1}
\issuenum{1}
\articlenumber{0}
\pubyear{2024}
\copyrightyear{2024}
\externaleditor{\textls[-25]{Academic Editor: Firstname Lastname}}
\datereceived{25 November 2024} 
\daterevised{11 December 2024} 
\dateaccepted{13 December 2024} 
\datepublished{ } 
\hreflink{\textls[-15]{https://doi.org/}} 
\pdfoutput=1

\usepackage{graphicx}
\usepackage{color}
\usepackage{amssymb}
\usepackage{upgreek}
\usepackage{bbm}
\usepackage{multirow}
\usepackage{multicol}
\usepackage{amscd}
\usepackage{ipa}
\usepackage{mfpic}
\usepackage{verbatim}
\usepackage{cancel}
\usepackage{chemarrow}
\usepackage{tikz-cd}
\usepackage{tocloft}
\usepackage{MnSymbol} 
\usepackage{upgreek}

\definecolor{cream}{rgb}{.97, .95, .88}
\definecolor{darkcream}{rgb}{1., .88, .5}
\definecolor{lightpink}{rgb}{0.98, 0.88, 0.87}
\definecolor{lightwhite}{rgb}{1., 0.98, 0.95}
\definecolor{lightsalmon}{rgb}{1., 0.95, 0.90}
\definecolor{lightviolet}{rgb}{0.9, 0.8, 0.9}
\definecolor{lightgray}{rgb}{.96, .96, .96}  
\definecolor{lgray}{rgb}{.75, .75, .75}
\definecolor{LemonChiffon}{rgb}{0.95, 1., 0.7}
\definecolor{lightolivegreen}{rgb}{0.84, 0.89, 0.25}
\definecolor{lightgreen}{rgb}{.664, 1., .52}
\definecolor{llgreen}{rgb}{.900, .983, .960}
\definecolor{tristle}{rgb}{0.87, 0.67, 0.77} 
\definecolor{pink}{rgb}{0.95, 0.45, 0.75}
\definecolor{magenta}{rgb}{1., 0, 1.}
\definecolor{violet}{rgb}{0.9, 0.20, 0.85}
\definecolor{darkolivegreen}{rgb}{0.55, 0.65, 0.35}
\definecolor{maroon}{rgb}{0.7, 0.26, 0.56}
\definecolor{lightmaroon}{rgb}{0.85, 0.38, 0.58}
\definecolor{darkmaroon}{rgb}{0.604, 0.169, 0.451}
\definecolor{ddarkmaroon}{rgb}{0.2, 0.03125, 0.150}
\definecolor{mediumorchid}{rgb}{0.8, 0.33, 0.83}
\definecolor{mediumorchidd}{rgb}{1., 0.33, 0.63}
\definecolor{darkgreen}{rgb}{0.1, 0.6, 0.13}
\definecolor{lightyellow}{rgb}{1., 1., 0.82}
\definecolor{turquoise}{rgb}{0.042, 0.586, 0.512}
\definecolor{turquoisel}{rgb}{0.66, 0.94, 0.83}
\definecolor{darkturquoise}{rgb}{0.21, 0.55, 0.50}
\definecolor{coral}{rgb}{1., 0.6, 0.21}
\definecolor{lightorange}{rgb}{1., 0.88, 0.75}
\definecolor{orangered}{rgb}{1., 0.5, 0.}
\definecolor{orange}{rgb}{1., 0.65, 0.1}
\definecolor{orangel}{rgb}{1., .85, .3}
\definecolor{darkorange}{rgb}{0.875, 0.4, 0.204}
\definecolor{ddarkorange}{rgb}{.675, .218, .05}
\definecolor{bluesky}{rgb}{0.48, 0.53, 1.}
\definecolor{gold}{rgb}{1., 0.85, 0.25}
\definecolor{goldd}{rgb}{0.95, 0.75, 0.05}
\definecolor{darkviolet}{rgb}{0.54, 0.04, 0.84}
\definecolor{ddarkviolet}{rgb}{.382, .063, .657}
\definecolor{lightblue}{rgb}{0.30, 0.86, 0.89}
\definecolor{LightBlue}{rgb}{0.68, 0.85, 0.9}
\definecolor{lblue}{rgb}{0.78, 0.90, 0.95}
\definecolor{darkblue}{rgb}{.105, .308, .707}
\definecolor{lightmaroon}{rgb}{0.85, 0.38, 0.58}
\definecolor{darkmaroon}{rgb}{0.604, 0.169, 0.451}
\definecolor{darkpink}{rgb}{0.879, 0.020, 0.766}
\definecolor{ddarkpink}{rgb}{0.738, 0.195, 0.406}
\definecolor{grey}{rgb}{0.717, 0.717, 0.717}
\definecolor{lightgrey}{rgb}{0.800, 0.800, 0.800}
\definecolor{brown}{rgb}{0.740, 0.323, 0.182}
\definecolor{redbrown}{rgb}{.575, .158, .05}
\definecolor{darkbrown}{rgb}{0.34, 0.25, 0.05}
\definecolor{orangebrown}{rgb}{0.433, 0.262, 0.06}
\definecolor{pinkl}{rgb}{1., 0.788, 0.918}
\definecolor{salmon}{rgb}{1., 0.66, 0.5}
\definecolor{lightbrown}{rgb}{0.703, 0.508, 0.121}

\def\Journal#1#2#3#4{{#1} {\bf #2}, (#3) #4}

\def\etal{{\it et al.}}
\def\Name#1#2 {{ #1 }{#2}}

\def\AA{\em A.\& A.}

\def\APN{\em Annalen Phys.}

\def\ATM{\em Adv. Theor. Math. Phys}

\def\CMP{\em Commun. Math. Phys.}

\def\CQG{\em Class. Quant.~Grav.}

\def\FOP{\em Found. Phys.}

\def\IMA{{\em Int. J. Mod. Phys.} \emph{A}}

\def\JHA{\em J. High Energy Astrophys.}

\def\JPA{\em J. Phys. A}
\def\JPC{\em J. Phys. Conf. Ser.}

\def\JPU{\em J. Phys. (USSR)}

\def\MDU{\em MDPI Universe J.}

\def\NAT{\em Nature}

\def\NJP{\em New J. Phys.}

\def\PLA{{\em Phys. Lett.}~\emph{A}}
\def\PLB{{\em Phys. Lett.}~\emph{B}}

\def\PRA{{\em Phys.~Rev.}~\emph{A}}

\def\PRD{{\em Phys.~Rev.}~\emph{D}}
\def\PRE{\em Phys.~Rep.}
\def\PRL{\em Phys. Rev. Lett.}

\def\PRV{\em Phys.~Rev.}
\def\PRX{{\em Phys.~Rev.}~\emph{X}}

\def\SYM{\em Symmetry}

\def\st{\scriptstyle}

\def\be{\begin{equation}}
\def\ee{\end{equation}}
\def\bea{\begin{eqnarray}}
\def\eea{\end{eqnarray}}
\def\bes{\begin{equation*}}
\def\ees{\end{equation*}}
\def\beas{\begin{eqnarray*}}
\def\eeas{\end{eqnarray*}}

\def\tr{\text{tr}}

\def\um{\mathcal U}

\def\hA{\hat{A}}

\def\hH{\hat{H}}
\def\hJ{\hat{J}}
\def\hL{\hat{L}}
\def\hO{\hat{O}}

\def\hT{\hat{T}}

\def\hrho{\hat{\rho}}
\def\hvarrho{\hat{\varrho}}

\def\sD{\cancel{D}}

\def\sqgr{$SU(\infty)$-QGR}
\def\suinf{{SU(\infty)}}
\def\suinfa{{\mathcal{SU}(\infty)}}

\Title{$\suinf$ Quantum Gravity and Cosmology}

\TitleCitation{$\suinf$ Quantum Gravity and Cosmology}

\Author{Houri Ziaeepour $^{1,2}$\orcidA{}}

\AuthorNames{Houri Ziaeepour}

\AuthorCitation{Ziaeepour, H.}

\address{%
$^{1}$ \quad 
Observatoire de Besan\c{c}on, Universit\'e de 
Franche Compt\'e, 41 bis Ave. de l'Observatoire, BP 1615, 25010 Besan\c{c}on, France; 
houriziaeepour@gmail.com\\
$^{2}$ \quad Mullard Space Science Laboratory, University College London, Holmbury St. Mary, 
GU5 6NT Dorking, Surrey, UK}

\newcounter{propos}

\abstract{
In this letter, we highlight the structure and main properties of an abstract approach to 
quantum cosmology and gravity, dubbed \sqgr. Beginning from the concept of the Universe 
as an isolated quantum system, the main axiom of the model is the existence of an infinite 
number of mutually commuting observables. Consequently, the Hilbert space of the Universe 
represents $\suinf$ symmetry. This Universe as a whole is static and topological. Nonetheless, 
quantum fluctuations induce local clustering in its quantum state and divide it into 
approximately isolated subsystems representing $G \times \suinf$, where $G$ is a generic 
finite-rank internal symmetry. Due to the global $\suinf$ each subsystem is entangled to the 
rest of the Universe. In addition to parameters characterizing the representation of $G$, 
quantum states of subsystems depend on four continuous parameters: two of them characterize 
the representation of $\suinf$, a dimensionful parameter arises from the possibility of 
comparing representations of $\suinf$ by different subsystems, and the fourth parameter 
is a measurable used as time registered by an arbitrary subsystem chosen as a quantum 
clock. It introduces a relative dynamics for subsystems, formulated by a symmetry-invariant 
effective Lagrangian defined on the (3+1)D space of the continuous parameters. At lowest 
quantum order, the Lagrangian is a Yang--Mills field theory for both $\suinf$ and internal 
symmetries. We identify the common $\suinf$ symmetry and its interaction with gravity. 
Consequently, \sqgr~predicts a spin-1 mediator for quantum gravity (QGR). Apparently, 
this is in contradiction with classical gravity. Nonetheless, we show that an observer 
who is unable to detect the quantumness of gravity perceives its effect as curvature 
of the space of average values of the continuous parameters. We demonstrate Lorentzian 
geometry of this emergent classical spacetime.
}


\begin{document}

\section{Introduction}  \label{sec:intro}
In the absence of a satisfactory quantum model for spacetime, gravity, and their 
relationship with other fundamental interactions, the model called \sqgr~takes an abstract 
approach to QGR. 

Since the first attempts to quantize gravity, model makers have tried to quantize either one 
of the various formulations of the Einstein gravity, such as the Arnowitt, Deser, and Misner 
(ADM) canonical (3+1)D formulation~\cite{admgr} and Loop Quantum Gravity 
(LQG)~\cite{ashtekarvar}, or another classical model in which a spin-2 field ---presumably a 
graviton field--- emerges. Such a field is usually associated to the 
Riemannian geometry of a spacetime, which sometimes is extended to include matter fields. 
The most popular version of such models is (super)string theory and its many variants. 
Despite many differences between these approaches, they have at least one common feature: 
the space (time) and other ingredients of these models should a priori persist if the 
Planck constant $\hbar \rightarrow 0$. However, in some cases, such as in LQG, the 
existence of a classical limit is not certain. Indeed, it is very difficult to prove the 
existence of a classical limit for the non-perturbative canonical formulation of LQG. Even 
in its perturbative form, that is, the spin foam formulation, finding a semi-classical 
limit is not straightforward or unique---see, 
e.g.,~\cite{spinfoamlimit,spinfoamlimit0,spinfoamlimit1}. In any case, decades of 
investigation into these models show that they suffer from various other issues, such as 
non-renormalizability; the lack of a genuine and fundamental relation between spacetime 
and matter (the source of gravitation); the prediction of Lorentz symmetry violation and 
modified dispersion relation, which are both stirringly 
constrained~\cite{grestgrb090510a,lorentzviolconstr}; the presence of extra dimensions that 
need compactification and leads to an exponentially large number of possible vacua and 
models; and finally, the lack of a convincing explanation for the observed (3+1)D classical 
spacetime. These issues have encouraged the suggestion of an emergent spacetime---see, 
e.g.,~\cite{spacetimeemerge}---and even the suggestion of a fundamentally classical nature 
for gravity~\cite{grclassic}.

On the other hand, $\hbar \rightarrow 0$ does not correspond to our Universe. Recent quantum 
technological achievements have shown that quantum mechanics should be applied to all scales 
and that classicality is an approximation. The evidence of macroscopic quantum effects is 
everywhere: the realization of quantum entanglement between billions of 
atoms~\cite{qmentanglemb}; the remote entanglement of micro-mechanical 
oscillators~\cite{qmentanglemm}; the crucial role of quantum mechanics in 
explaining exotic phenomena in condensed matter, such as superconductivity, superfluidity, 
and topological effects (not only in a laboratory setting, but also in astronomical 
objects such as giant planets, white dwarfs, and neutron stars); and the necessity of a 
quantum-generated inflation to explain cosmological observations. 

Considering the difficulties in standard quantization approaches to QGR and the prominence 
of quantum mechanics in the fabric of the Universe, it seems reasonable to seek a 
fundamentally quantum approach to cosmology and QGR. The \sqgr~is not the first model in 
this category ---called Quantum First Models (QFM) by some 
authors~\cite{qgrlocalqm1}. Nonetheless, it is significantly different from QFM that try 
to introduce an intrinsic concept of locality in Quantum Field Theory (QFT)--- assumed 
to be important for the physics of black holes~\cite{qgrhistory,qgrlocalqm0,qgrlocalqm2}. 
It also significantly diverges from models that use information, graph theory and/or 
quantum tensor networks in place of a spacetime, and interpret gravity as their 
entanglement~\cite{qgrentangle2,qgrentangle3,qgrentangle1} (and references therein). 
A main topic that is crucial for the introduction of structure, locality, matter, and 
internal symmetries in the QFM is the concept of subsystem. This issue is not addressed 
in a systematic way in the previous models. One of the goals of \sqgr~is to introduce and 
study this \mbox{fundamental concept.}

The \sqgr~was first reported in~\cite{houriqmsymmgr} and in more detail in~\cite{hourisqgr}. 
Some of the technical details and demonstrations are reported in~\cite{hourisqgrym}, and 
the model is compared with some of other approaches to quantum gravity, including QFM 
mentioned above in~\cite{hourisqgrcomp}. The aim of this letter is to provide a concise 
description of the essential features of the model as studied so far. In-depth explanations 
of the reviewed topics and their mathematical derivation can be found 
in~\cite{houriqmsymmgr,hourisqgr,hourisqgrym,hourisqgrcomp}. 

In Section~\ref{sec:axiom}, we briefly describe the axioms of \sqgr~and how they lead to a 
Hilbert space for the Universe, representing $\suinf$ symmetry. Section~\ref{sec:symm} 
reviews the properties of this Hilbert space and its vectors as a quantum state of the 
Universe. In Section~\ref{sec:dynuniv}, we define a symmetry-invariant functional on 
the 2D space of parameters that characterize the representation of $\suinf$ by the Hilbert 
space. We show that, as a Lagrangian for the whole Universe, this functional is static and 
the only difference between possible universes is the topology of their parameter space. 
In Section~\ref{sec:subsys}, we argue that quantum fluctuations can locally ---in the 
Hilbert space--- break the symmetry of states and induce structures representing 
generic finite-rank symmetries $G$. Consequently, according to well-defined criteria for 
the division of a quantum system to subsystems~\cite{sysdiv}, these structures can be 
interpreted as approximately isolated subsystems. Indeed, in Section~\ref{sec:entangle}, 
we show that, due to the global $\suinf$ symmetry, these subsystems are not separable and 
each of them is quantum-entangled to the rest of the Universe 
(Proposition \ref{propentang}). In Section~\ref{sec:subsyssymm}, we demonstrate that, due 
to this global entanglement, the Hilbert spaces of subsystems represent $G \times \suinf$. 
As $\suinf$ symmetry is shared by all subsystems, the corresponding interaction can be 
identified with gravity. The space of parameters characterizing the states 
of subsystems and their relative dynamics is discussed in Section~\ref{sec:param}. We 
demonstrate that they include four continuous parameters. But the geometry of their space 
$\Xi$ is arbitrary and its reparameterization is equivalent to a $\suinf$ gauge 
transformation, under which subsystems should be invariant (Proposition 
\ref{propparamcurve}). Hence, despite apparent similarity to the classical spacetime, 
$\Xi$ cannot be identified with it. Nonetheless, in Section~\ref{sec:geom}, we use Quantum 
Speed Limits (QSLs), originating from quantum uncertainty relations, to define the 
average/expectation values of parameters and an average path in the parameter space for 
the subsystems. These average/expectation quantities define a classical space $\Sigma$, 
and we demonstrate that its geometry is Lorentzian and related to the quantum state of 
subsystems. We identify this emerged (3+1)-dimensional space with the perceived classical 
spacetime. In Section~\ref{sec:evol}, we 
construct a symmetry-invariant effective Lagrangian for subsystems on their (3+1)D space 
of continuous parameters $\Xi$ and demonstrate that, at lowest quantum order, it has the 
form of a Yang--Mills theory for both $G$ and $\suinf$ symmetries. Finally, in 
Section~\ref{sec:limit}, after clarifying the meaning of a classical limit in \sqgr, 
we show that, in this limit, its Lagrangian is similar to a QFT in curved spacetime with 
Einstein--Hilbert action for gravity. Section~\ref{sec:conclu} includes concluding remarks. 

\section{Axioms of \boldmath{\sqgr}}  \label{sec:axiom}
The construction of \sqgr~begins with considering the Universe as an isolated quantum system 
satisfying the following rules and properties:
\setcounter{enumi}{0}
\renewcommand{\theenumi}{\Roman{enumi}}
\begin{enumerate}[leftmargin=22pt,labelsep=1pt]
\item Quantum mechanics is valid at all scales and applies to every entity, including the 
Universe as a whole;\label{uniaxiom1}
\item Every quantum system is exclusively described by its symmetries, and its Hilbert space 
represents them;\label{uniaxiom2}
\item The Universe has an infinite number of independent degrees of freedom associated to as 
many mutually commuting quantum observables.\label{uniaxiom3} 
\end{enumerate}

Examples of macroscopic quantum systems and effects discussed in the Introduction justify 
the axiom \ref{uniaxiom1}. As for the axiom \ref{uniaxiom2}, the postulates of quantum 
mechanics, \`a la Dirac and von Neumann, do not specify how the Hilbert space should be 
chosen. Nonetheless, in practice, it always represents the symmetries of the quantum system. 
Indeed, it is possible to reformulate quantum mechanics postulates with symmetry as a 
foundational concept~\cite{houriqmsymm}. Finally, the huge number of observed approximately 
separable elementary particles, each having a number of independent quantum observables, 
motivates the axiom \ref{uniaxiom3}.

Considering the condition of the hermitianity of operators associated to quantum 
observables and the unitarity of basis transformation, the Hilbert space ${\mathcal H}_U$ 
and the space of (bounded) linear operators ${\mathcal B}[{\mathcal H}_U]$ are 
infinite-dimensional and represent $\suinf$ symmetry. It is shown~\cite{suninfrep0} that 
all simple compact Lie groups converge to $\suinf$ when their rank $N \rightarrow \infty$. 
Therefore, $\suinf$, as the symmetry of such a Universe, is unique. 

\section{Hilbert Space of the Universe and Quantization}  \label{sec:symm}
Representations of $\suinf$ are homomorphic to area-preserving diffeomorphisms of 2D compact 
Riemann surfaces $D_2$; their associated algebra is homomorphic to that of the Poisson 
brackets~\cite{suninfhoppthesis,suninftorus,suninfsurfaceanomal,suinftriang,suinftriang0,suninfrep} 
(see the appendices in~\cite{houriqmsymmgr,hourisqgr,hourisqgrym} for a review of these 
representations, other properties of $\suinf$ and its algebra, and more references). Thus, 
${\mathcal B}[{\mathcal H}_U] \cong SU(\infty) \cong ADiff(D_2)$, where, throughout this 
work, the symbol $\cong$ means ``homomorphic to'', and ${\mathcal B}[{\mathcal H}_U]$ 
is the space of (bounded) linear operators acting on ${\mathcal H}_U$. We call the 2D 
compact surface associated to a representation of $\suinf$ its ``diffeo-surface''. 

Generators of $\suinf$ have the following general expression~\cite{suinftriang0}:
\be
\hL_f = \frac{\partial f}{\partial \eta} \frac{\partial }{\partial \zeta} - 
\frac{\partial f}{\partial \zeta} \frac{\partial }{\partial \eta} \quad , 
\quad \hL_f ~ g = \{f,g\}, \quad [\hL_f, \hL_g] \equiv \hL_{\{f,g\}} \label{suinfgendef}
\ee
where $f$ and $g$ are any $C^\infty$ scalar function on the diffeo-surface $D_2$ and 
$(\eta, \zeta)$ are local coordinates on the surface. It is more convenient to use the  
decomposition of functions $f$ and $g$ to orthonormal functions. Two popular decompositions, 
called ``spherical'' and ``torus'' bases, use spherical harmonic~\cite{suninfhoppthesis} 
and 2D Fourier transform~\cite{suninftorus,suinftriang,suinftriang0}, respectively. We 
indicate the decomposed generators of $SU(\infty)$ as $\hL_a (\eta, \zeta)$. In sphere basis, 
$a \equiv (l, m),~ l \in \mathbb{Z^+},~ -l \leqslant m \leqslant l$ and, in torus basis, 
$a \equiv (m,~n),~ m, n \in \mathbb{Z}$. 

In the absence of a background spacetime in \sqgr,~the non-Abelian algebra in (\ref{suinfgendef}) 
replaces the usual quantization relations~\cite{qmmathbook,qgrnoncommut}. Nonetheless, as 
this algebra has an infinite rank and is characterized by two continuous variables, it is 
also possible to find conjugate operators $\hJ_a$ for $\hL_a$ such that 
$[\hJ_a,\hL_b] = -i \hbar \delta_{ab} \mathbbm{1}$, where 
$\hJ_a \in {\mathcal B}[{\mathcal H}_U^*]$ and${\mathcal H}_U^*$ is the dual Hilbert space. 
See the appendices in~\cite{hourisqgr} for the decomposition of $\hL_a$ and $\hJ_a$, and 
their analogy with position and momentum operators in models with a background spacetime 
according to the Weyl quantization scheme. Notice that, in the above commutation relation, 
we have explicitly shown the Planck constant $\hbar$. Indeed, as the generators of the 
Hilbert space of a quantum system, operators $\hL_f$ and $\hL_g$ in (\ref{suinfgendef}), and 
their homologous $\hL_a$ in a specific basis for $\suinf$ algebra, should be normalized such 
that the r.h.s. of the commutation relation becomes proportional to $\hbar$. Moreover, 
later in this work, we show that a dimensionful constant, which can be chosen to be the 
Planck mass $M_P$ or the Planck length $L_P \propto 1/M_P$, arises in the model when the 
Universe is divided into subsystems. This constant can be included in the generators such 
that the r.h.s. of (\ref{suinfgendef}) becomes proportional to $\hbar/M_P$. This 
normalization corresponds to using $\suinf$ gauge fields, defined in (\ref{yminvardef}), as 
generators of the $\suinfa$ algebra and shows that, for $\hbar \rightarrow 0$ or 
$M_P \rightarrow \infty$ ---corresponding to classical limit or no gravity, 
respectively--- the algebra in (\ref{suinfgendef}) becomes Abelian and its associated 
symmetry group $\otimes^{N \rightarrow \infty} U(1) \cong \otimes^{N \rightarrow \infty} \mathbb{R}$. 
This is the symmetry of a classical system with $N \rightarrow \infty$ independent 
observables. However, in this case, the diffeo-surface as a parameter space for the states 
of the system would not be present. Thus, to have a meaningful quantum model, $\hbar/M_P$ 
must remain finite. In other words, according to \sqgr,~quantumness and gravity 
are~inseparable.

\section{A Globally Static and Topological Universe}  \label{sec:dynuniv}
This isolated Universe is, by construction, static because there is no external system 
that can be used as a clock. Indeed, it is a trivial quantum system because every state 
vector $\psi \in {\mathcal H}_U$ can be transformed to another state by a unitary 
transformation $U \in {\mathcal B}[{\mathcal H}_U] \cong \suinf$. Such transformations can 
be considered as changes in the Hilbert space basis. But, as there is no preferred (pointer) 
basis, all bases and states are physically equivalent. The triviality of the model can be 
also verified by considering an effective Lagrangian functional ${\mathcal L}_U$ that is 
invariant under $\suinf$~\cite{houriqmsymmgr}. Such a functional can be constructed from 
the traces of the products of symmetry generators. In analogy with QFT, we only consider 
the lowest-order traces because higher order functionals can be constructed through path 
integral formalism. The application of the variational principle shows that the solution 
of dynamical equations of the ``fields'' ---the coefficients of trace 
terms--- is locally trivial, but unstable under fluctuations~\cite{houriqmsymmgr}. This 
means that states can locally ---in the Hilbert space--- become clustered. We 
discuss this process in more detail in the following~sections. 

Using the logical requirement that the Lagrangian should be invariant under the 
reparameterization of the diffeo-surface of $\suinf$, we find~\cite{hourisqgr} that, at 
lowest order in traces, ${\mathcal L}_U$ has an expression similar to a 2D Yang--Mills theory on the diffeo surface $D_2$:
\bea
& {\mathcal L}_U = \upkappa \int d^2\Omega ~ \biggl [ ~\frac{1}{2} ~ \tr (F^{\mu\nu} F_{\mu\nu}) + 
\frac {1}{2} \tr (\sD \hrho_U) \biggr ], \quad \quad \mu, \nu \in {\theta, \phi} 
\label{yminvar} \\
& F_{\mu\nu} \equiv F_{\mu\nu}^a \hL^a \equiv [D_\mu, D_\nu], \quad 
D_\mu = (\partial_\mu - \Gamma_\mu) \mathbbm{1} + \sum_a i \lambda A_\mu^a \hL^a, 
\quad a \equiv (l,m) \label{yminvardef} \\
& F_{\mu\nu}^a F^{\mu\nu}_a = L^*_a L^a, \quad \forall a. \label{ltof} 
\eea
where as an example index $a$ is expanded for sphere basis. The first term in (\ref{yminvar}) 
does not depend on the geometric connection of $D_2$. Indeed, 
it is well-known that, in any dimension, only the covariant derivatives of Yang--Mills gauge 
fields depend on the geometry of background space. In QFT on an independent background, the 
second term is not topological. However, in ${\mathcal L}_U$, the space $D_2$ is the 
diffeo-surface of its $\suinf$ symmetry. Every deformation of $D_2$ can be decomposed to an 
area-preserving deformation and a global scaling that changes the irrelevant constant 
$\upkappa$. Such operation rescales the area of $D_2$, but the area of the diffeo-surface 
is irrelevant for its relationship with $\suinf$ and its representation---see Diagrams 
(\ref{singlesuinfdiag}) and (\ref{multisuinfdiag}) below. Therefore, a change in $D_2$ 
geometry can be neutralized by an $\suinf$ transformation;  see~\cite{hourisqgrym} for a 
detailed demonstration. Consequently, ${\mathcal L}_U$ does not depend on the local geometry 
of the diffeo-surface $D_2$ and can be considered as a topological action.

The 2D Riemannian surfaces $D_2$ are topologically classified by the Euler characteristic 
$\chi (D_2)$:
\be
\int d^2\Omega ~ \mathcal{R}^{(2)} = 4\pi ~ \chi (D_2), \quad\quad \chi (D_2) = 
2 - {\mathcal G} (D_2), \quad \quad \chi (D_2) = 2 - {\mathcal G} (D_2) \label{eulercharct}
\ee
where $\mathcal{R}^{(2)}$ is the scalar curvature of $D_2$ and ${\mathcal G}$ is its genus. 
The topological nature of the Lagrangian in (\ref{yminvar}) ---in particular, the pure 
gauge term--- implies that, without any loss of generality, its integrand must be 
proportional to $\mathcal{R}^{(2)}$:
\be
\tr (F^{\mu\nu} F_{\mu\nu}) ~ \propto ~ \mathcal{R}^{(2)}   \label{gaugecurve}
\ee
This relation becomes an equality by changing the arbitrary normalization of $\suinf$ 
gauge fields. In addition, Equation (\ref{gaugecurve}) is the first hint about the 
relationship between the action ${\mathcal L}_U$ with gravity because, in contrast to 
2D Yang--Mills on a classical background, the space $D_2$, on which this action is defined, 
is related to the Yang--Mills gauge symmetry itself.

\section{Emergence of Structures and Subsystems}  \label{sec:subsys}
The assumed global $\suinf$ symmetry prevents an exact division of the Universe into 
separable subsystems according to the criteria defined in~\cite{prodstate}. Nonetheless, 
here, we show that its infinite number of observables ---degrees of freedom--- are 
sufficient for an approximate blockization of states and a description of ${\mathcal H}_U$ 
as a tensor product. 
 
A divisible quantum system must fulfill specific conditions~\cite{sysdiv}. In particular, 
linear operators applied to its state should consist of mutually commuting subsets 
$\{\hA_i\}$s, where each subset represents an internal symmetry $G_i$. Another way 
to distinguish subsystems in a quantum system is through the factorization of its state. 
In~\cite{hourisqgrym}, we show that these two definitions are equivalent.  

At this stage, there is no concept of time in \sqgr~and, therefore, no ``order of events''. 
For this reason, to show how approximately isolated subsystems arise in the 
\sqgr~Universe, we use an operational approach. Consider the Universe in a completely 
coherent state, defined as $\hrho^{CC} \equiv {\mathcal N} \sum_{a,b} |a\rangle \langle b|$ 
in an arbitrary basis ,where ${\mathcal N}$ is a normalization constant. By definition, 
this state has maximum coherence according to any of the coherence measures suggested 
in~\cite{qminfocohere}. Thus, the application of a unitary operator to it transforms 
$\hrho^{CC}$ to a less coherent state. For instance, using the sphere basis for $\suinf$ 
representation, a general state can be written as
\be
|\psi_U\rangle = \int d\Omega \sum_{\substack{l \geqslant 0 ,\\ -l \leqslant m \leqslant l}} \psi_U^{lm} 
|{\mathcal Y}_{lm} (\theta, \phi)\rangle, \quad \quad |{\mathcal Y}_{lm} (\theta, \phi)\rangle 
\equiv Y_{lm} (\theta, \phi) |\theta, \phi \rangle  \label{ystateexpan}
\ee
In $|{\mathcal Y}_{lm} (\theta, \phi)\rangle$ basis, the completely coherent state corresponds 
to $\psi_U^{lm} = const.$ Assume that quantum fluctuations lead to the application of 
$\hL_{l_1m_1}$ on $|\psi^{cc}\rangle$ and changes it to
\bea
\int d\Omega ~ \hL_{l_1m_1}(\theta, \phi) |\psi^{cc} \rangle & = & {\mathcal N} \int d\Omega 
\sum_{\substack{l \geqslant 0 ,\\ -l \leqslant m \leqslant l}} \hL_{l_1m_1} (\theta, \phi) 
|{\mathcal Y}_{lm} (\theta, \phi)\rangle \nonumber \\
& = & -i \hbar {\mathcal N} \int d\Omega \sum_{\substack{l \geqslant 0 , -l \leqslant m \leqslant l \\ 
l' \geqslant 0 , -l' \leqslant m' \leqslant l'}} f ^{l'm'}_{l_1 m_1,lm} Y_{l'm'} (\theta, \phi) 
|\theta, \phi \rangle  \nonumber \\
& = & -i \hbar {\mathcal N} \int d\Omega \sum_{\substack{l \geqslant 0 , -l \leqslant m \leqslant l \\ 
l' \geqslant 0 , -l' \leqslant m' \leqslant l'}} f ^{l'm'}_{l_1 m_1,lm} 
|{\mathcal Y}_{l'm'} (\theta, \phi) \rangle  \equiv |{\mathbf g}_{ l_1 m_1} \rangle  
\label{opstateexpand}
\eea
where the $\suinf$ structure constants $f ^{l'm'}_{l_1 m_1,lm}$ are proportional to 3j symbols 
and depend on the indices $(l,m), (l',m'), (l'_1,m'_1)$~\cite{suninfhoppthesis}. In general, 
the new state $|{\mathbf g}_{l_1 m_1}(\theta, \phi)\rangle$ and its corresponding density 
matrix are not completely coherent anymore, but more structured. Specifically, using the 
norm of the off-diagonal components of the density matrices of $|\psi_U\rangle$ and 
$|{\mathbf g}_{ l_1 m_1} \rangle$ as a measure of coherence~\cite{qminfocohere} and the 
boundedness of the integrals of spherical harmonic functions~\cite{spherharmonbound}, on 
which the coefficients $f ^{l'm'}_{l_1 m_1,lm}$ depend, it is straightforward to show that 
$|\psi_U\rangle$ is maximally coherent, but $|{\mathbf g}_{ l_1 m_1} \rangle$ is much less 
so ---in other words, it is more clustered/blockized (details of this calculation will 
be reported elsewhere). Moreover, the blockization of the density matrix is more probable 
to grow with the successive application of $\hL_a$s because there are an infinite number 
of $\hL_a$ operators with $a \equiv (l,m) \neq (l_1, m_1)$, and the probability of a 
random occurrence of $\hL_{l_1m_1}^{-1}$ after operation (\ref{opstateexpand}) is extremely 
small. Of course, as we explained earlier, all these states are globally equivalent and 
can be transformed to each others without changing physical observables. But, locally and 
approximately, these blocks satisfy the subsystem criteria defined in~\cite{sysdiv}. 
Therefore, ${\mathcal H}_U$ and ${\mathcal B}[{\mathcal H}_U]$ can be approximately 
decomposed as:
\bea
&& {\mathcal H}_U \leadsto \bigoplus_i {\mathcal H}_i \leadsto \bigotimes_i {\mathcal H}_i  
\label{hilbertmulti} \\
&& {\mathcal B}[{\mathcal H}_U] \leadsto \bigoplus_i {\mathcal B}[{\mathcal H}_i] \leadsto 
\bigotimes_i {\mathcal B}[{\mathcal H}_i]  \label{hilbertsum}
\eea
where the symbol $\leadsto$ means ``approximately leads to''. To demonstrate the 
emergence of this approximate tensor product more explicitly and in a 
representation-independent manner, we use the properties of the Cartan decomposition of 
$\suinf$. Specifically, $\suinf$ can be decomposed to an infinite tensor product of any 
finite-rank Lie group~\cite{suninfrep0}:
\bea
\suinf & \cong & \bigotimes_i^\infty G_i  \label{gdecomp} \\
\suinf^n & \cong & \suinf ~ \forall n  \label{suinfgdecomp}
\eea
where $G_i$s are finite-rank Lie groups and can be different from each others. Moreover, 
$G_i$'s themselves can be the products of groups with smaller ranks. However, here, for the 
sake of the simplicity of notation, we generically call them $G$.  

Considering (\ref{gdecomp}) and (\ref{suinfgdecomp}), the Hilbert space ${\mathcal H}_U$ 
can be decomposed to
\be
G \times \suinf \cong \suinf  \label{gsuinfcong}
\ee
Notice that, if $G$ had an infinite rank, the subsystem would be indistinguishable 
from the whole Universe. Moreover, Equations~(\ref{gdecomp}) and (\ref{suinfgdecomp}) can 
be combined to
\be
\suinf \cong \bigotimes^\infty (G \times \suinf)  \label{gsuinf}
\ee
Following these decompositions, the state of the Universe can be written as a tensor product: 
\bea
|\Psi_U\rangle & = & \sumint_{(\eta, \zeta,)} A_U (\eta, \zeta) |\psi_U(\eta, \zeta) \rangle = 
\sumint_{\{k_G\}, \{y\}} A (k_G; y) ~ |\psi_G(k_G) \rangle \times |\psi_\infty(y) \rangle, 
\nonumber \\
&& y \equiv (\eta, \zeta; \cdots) \label{gustate}
\eea
and its corresponding density matrix as

\bea
\hvarrho_U & = & \sumint_{(\eta, \zeta, \eta', \zeta')} A_U (\eta, \zeta)  
A_U^*(\eta', \zeta') ~ \hrho_U (\eta, \zeta, \eta', \zeta')  \nonumber \\
& = & \sumint_{\substack {\{k_G, k'_G\} \\ \{y, y'\}}} A (k_G; y)  A^*(k'_G, y') ~ 
\hrho_G (k_G , k'_G) \times \hrho_\infty (y, y') \label{gudensity} 
\eea
\bea
& \hrho_U (\eta, \zeta, \eta', \zeta') \equiv |\psi_U (\eta, \zeta) \rangle 
\langle \psi_U (\eta', \zeta')|, \label{udensitydef} \\
& \hrho_G (k_G , k'_G) \equiv |\psi_G (k_G) \rangle \langle \psi_G (k'_G)|, \quad 
\hrho_\infty (y, y') \equiv |\psi_\infty (y) \rangle \langle \psi_\infty (y')|  
\label{gudensitydef}  \\
& \sumint_{(\eta, \zeta)} |A_U (\eta, \zeta)|^2 = 1, \quad \quad 
\sumint_{\substack {\{k_G\} \\ \{y\}}} |A (k_G; y)|^2 = 1  
\label {amptracecond} 
\eea
The bases $\{|\psi_U(\eta, \zeta)\}$, $\{|\psi_G(k_G) \rangle\}$, and $\{|\psi_\infty(y)\}$ 
generate the Hilbert spaces of the Universe ${\mathcal H}_U$, and subspaces 
${\mathcal H}_G \subset {\mathcal H}_U$ and ${\mathcal H}_{\infty} \subseteq {\mathcal H}_U$ 
that represent $G$ and $\suinf$, respectively. Accordingly, operators 
$\hrho_G (k_G , k'_G) \in {\mathcal B}[{\mathcal H}_G]$ and 
$\hrho_\infty(y) \in {\mathcal B}[{\mathcal H}_{\infty}]$, such that 
\mbox{${\mathcal B}[{\mathcal H}_G] \times \mathbbm{1}_\infty \subset {\mathcal B}[{\mathcal H}_U]$} 
and $\mathbbm{1}_G \times {\mathcal B}[{\mathcal H}_\infty] \subset {\mathcal B}[{\mathcal H}_U]$ 
are the bases of ${\mathcal B}[{\mathcal H}_G]$ and ${\mathcal B}[{\mathcal H}_\infty]$, 
respectively. The operator set $\{\hrho_U (\eta, \zeta, \eta', \zeta')\}$ is a basis for 
${\mathcal B}[{\mathcal H}_U]$. The set $\{k_G\}$ parameterizes the representation of $G$. 
For finite-rank Lie groups, the number of independent $k_G$s, that is the dimension $d_G$ of 
the parameter space $\{k_G\}$, is finite and $k_G$s usually take discrete values. 
For example, for $G= SU(2)$, $k_G = (l, m), ~ l \in \mathbb{Z}^+, ~ -l \leqslant m \leqslant l$. 
For a fixed $l$ ---corresponding to a super-selected representation of 
$SU(2)$--- the dimension $d_G = 2l +1$. The continuous parameters $(\eta, \zeta)$ are 
the coordinates of the diffeo-surface and characterize the generators of $\suinf$. 
The extension dots in $y$ and $y'$ indicate emergent parameters when the Universe is 
perceived through the ensemble of its subsystems. They are described in the following sections. 

It is easy to verify that ${\mathcal H}_G$ and ${\mathcal H}_\infty$ fulfill the requirements 
for subsystems defined in~\cite{sysdiv}. In a given basis for ${\mathcal H}_U$, they are, 
by construction, orthogonal to each other. Considering (\ref{gsuinfcong}), endomorphism 
condition ${\mathcal H}_U \cong {\mathcal H}_G \times {\mathcal H}_\infty$ is fulfilled. 
In the absence of a background spacetime, the locality condition in the usual sense is 
irrelevant. Nonetheless, due to the contractivity of distance functions~\cite{qminfobook}, 
the distance between states belonging to ${\mathcal H}_G$ is always smaller than their 
distance from similar states with non-zero projection in the complementary subspace 
${\mathcal H}_\infty $, and vice versa. Hence, the decomposition in (\ref{gudensity}) 
induces a ``locality'' concept and structure in the Hilbert space ${\mathcal H}_U$. 
This is in addition to the geometrical locality, which can be defined for any Hilbert space 
by associating a Fubini--Study metric and distance to its states.

It is important to notice that the concept of subsystem is more general than elementary 
particles or fields, which can be defined as a subsystem that cannot be decomposed to smaller 
parts. By contrast, a general subsystem can include many or even an infinite number of 
elementary subsystems. Similar to classical gravity, the \sqgr~formulation is independent 
of internal symmetries and the properties of subsystems. Therefore, it fully respects 
the equivalence principle.

\section{Global Entanglement}  \label{sec:entangle}
The difference between the Hilbert space of the Universe ${\mathcal H}_U$ that represents 
$\suinf$ and its states $|\psi_U \rangle$, and those of ${\mathcal H}_\infty $, is better 
understood if we use the properties of $SU(N \rightarrow \infty)$ Cartan decomposition and write 
(\ref{gsuinfcong}) as 
\be
SU(N \rightarrow \infty) \supseteq SU(K) \times SU(N-K \rightarrow \infty) \supseteq 
G \times SU(N-K \rightarrow \infty)  \label{suninfg}
\ee
in which $\infty > K \in \mathbb{Z}^+$ is chosen such that $SU(K) \supseteq G$. From this 
relation, it is clear that symmetry in the r.h.s. of (\ref{suninfg}) is smaller and presents a 
broken version of the l.h.s. Only when $N \rightarrow \infty$ are the two sides are 
homomorphic. Thus, only in this limit, 
${\mathcal H}_U \cong {\mathcal H}_G \times {\mathcal H}_\infty$. Otherwise, the factorization 
of $G$ from $SU(N \gg 1)$ due to the clustering of states presents an (approximate) breaking 
of the symmetry of the Universe. In the limit of $N \rightarrow \infty$, the Hilbert space 
${\mathcal H}_U$ can be decomposed to an infinite number of subsystems representing the 
generic finite-rank symmetry $G$. As discussed in Section~\ref{sec:subsys}, the finite-rank 
internal symmetry of subsystems do not need to be the same because 
$SU(N-K \rightarrow \infty)$ in (\ref{suninfg}) can be in turn decomposed, as long as the 
total rank of factorized groups $K' < \infty$ and the remaining $SU(N-K' \rightarrow \infty)$. 
Thus, Equations~(\ref{gdecomp})--(\ref{gsuinf}) and (\ref{suninfg}) show that the \sqgr~Universe can be 
constructed either top-down, that is, by dividing it into an infinite number of finite-rank 
subsystems (contents), or bottom-up, by considering it as the ensemble of an infinite number of 
quantum systems representing a symmetry, which can have any rank, including infinity. However, 
in the latter case, one should also impose the global $\suinf$ symmetry to connect everything 
together, as we show in the next sections. Thus, the top-down approach is more economical in 
the number of axioms. 

As we discussed earlier, despite the tensor product structure of the bases in (\ref{gustate}) 
and (\ref{gudensity}), and their corresponding Hilbert spaces, due to the global $\suinf$, 
amplitudes $A (k_G; \eta, \zeta, \cdots)$ are not factorizable. Consequently, $G$-representing 
subsystems are not separable ---they are entangled. This observation is formulated in the 
following proposition.

\refstepcounter{propos}

\begin{Proposition} \label{propentang}
 {In \sqgr,~every subsystem is entangled to the rest of the Universe.}
\end{Proposition}

In~\cite{hourisqgr}, mutual information is used to prove this proposition. We call this 
attribute of the model `the global entanglement''. A more explicit demonstration of 
Proposition \ref{propentang} consists of tracing out the $\suinf$-representing component of 
$\hvarrho_U$~\cite{hourisqgrym}:
\be
\hvarrho_G \equiv \tr_\infty \hvarrho_U = \int dy^D \sum_{\{k_G, k'_G\}} 
A_G (k_G; y) A^*_G (k'_G, y) ~ \hrho_G (k_G , k'_G), \quad \quad y \equiv (\eta, \zeta, \cdots) 
\label{gdensity}
\ee
It is shown that $\hvarrho_G$ is a mixed state and has a non-zero von Neumann 
entropy~\cite{hourisqgrym}. This result is not a surprise because, due to the global $\suinf$ 
symmetry, amplitudes $A_G (k_G; y)$ are not factorizable to $k_G$ and $y$ dependent functions. 
Therefore, ${\mathcal H}_G$ and ${\mathcal H}_\infty$ cannot be considered Hilbert spaces of 
separable subsystems. Nonetheless, the subspace ${\mathcal H}_G$ is approximately isolated by 
its ``local'' symmetry $G$. Moreover, considering the finite rank of $G$ and the 
entanglement of $\hvarrho_G$ with the rest of the Universe, it can be interpreted as the 
mixed state of a subsystem approximately isolated from its infinite-dimensional 
``environment'' due to their approximate inaccessibility,. Therefore, decompositions of the 
type presented in (\ref{suninfg}) induce a concept of "locality" for the subsystems. 
In addition, we notice that the amplitudes $A_G (k_G; y)$ have a structure similar to gauge 
fields, that is, they depend on the parameters of a finite-rank Lie group and a continuous 
``background''. Observables of the state $\hvarrho_G$ ---that is, Hermitian operators 
in ${\mathcal B}[{\mathcal H}_G]$--- are, by construction, invariant under application 
of $G$ and reparameterization of the "external parameters" $y$. We discuss the 
meaning and importance of these properties when relative dynamics for subsystems are 
introduced in Section~\ref{sec:evol}. 

Similarly, tracing out the $\hrho_G$ component of $\hvarrho_U$ leads to a mixed state 
$\hvarrho_\infty$:
\bea
&& \hvarrho_\infty \equiv \tr_G ~ \hvarrho_U  = 
\sumint_{\{(\eta, \zeta, \cdots)\}, \{(\eta', \zeta', \cdots)\}}  
{\mathcal A}_\infty (\eta, \zeta; \eta', \zeta', \cdots) ~ \hrho_\infty (\eta, \zeta ; \eta', \zeta', \cdots) 
\label{infdensity} \\
&& {\mathcal A}_\infty (\eta, \zeta; \eta', \zeta', \cdots) \equiv  
\sum_{\{k_G\}} A_\infty (k_G; \eta, \zeta, \cdots) A^*_\infty (k_G; \eta', \zeta', \cdots)  
\label{rhoinfamp} \\
&& \sumint_{\{(\eta, \zeta, \cdots)\}, \{(\eta', \zeta', \cdots)\}}  
{\mathcal A}_\infty (\eta, \zeta; \eta, \zeta, \cdots) = 1  \label {rhoinfamprenorm}
\eea
for the $\suinf$-representing environment. We notice that $\hvarrho_\infty$ also depends on 
a set of external parameters $\{k_G\}$, which are not related to the $\suinf$ symmetry of 
this state.. The physical meaning of this dependence is clarified once we establish, in 
Section~\ref{sec:geom}, an effective path for the subsystems in their parameter space.

The entanglement of mixed states $\hvarrho_G$ and $\hvarrho_\infty$ can be quantified using 
usual entanglement measures and are calculated in~\cite{hourisqgrym} for future applications of 
\sqgr. 

\section{The Full Symmetry of Subsystems}  \label{sec:subsyssymm}
In the mixed state $\hvarrho_G$, the parameter vector $y$ is in part the footprint of 
$\suinf$ symmetry and plays the role of a classical background for an observer who does 
not have access to the full extent of the quantum state of the Universe $\hvarrho_U$. 
On the other hand, considering axioms \ref{uniaxiom1} and  \ref{uniaxiom2} of \sqgr~about the  
direct or indirect quantum origin of all processes and observables, and their association to 
symmetries represented by the Hilbert space, the observer can associate two of the four 
components of $y$ ---we show this later--- to a representation of $\suinf$ symmetry and use 
them to purify $\hvarrho_G$ by extending the Hilbert space with an auxiliary space representing 
this symmetry. In~\cite{hourisqgrym}, we showed that $\hvarrho_G$ satisfies the conditions for 
faithful purification~\cite{qmpurifcond}. The purified state will have the 
following form:
\be
|\psi_{G_\infty}\rangle \equiv \sumint_{\{k_G\}; \{y\}} A_{G_\infty} (k_G; y) ~ |
\psi_G (k_G) \rangle \times |\psi_\infty (y) \rangle  \label{gpurified}
\ee
where $|\psi_\infty (y)$ has the same definition as in (\ref{gustate}), but is not necessarily 
the same basis. Although $|\psi_{G_\infty}\rangle$ looks like the state of the Universe 
$|\psi_U\rangle$ in (\ref{gustate}), according to the Schr\"odinger--HJW 
theorem~\cite{qmpurif,qmpurif0,qmpurif1} about the degeneracy of purification, in general, 
$|\psi_{G_\infty}\rangle \neq |\psi_U \rangle$. The state $|\psi_{G_\infty}\rangle$ can be also 
considered as a purification of $\hvarrho_\infty$. In both cases, the state is a vector in a 
Hilbert space that represents $G \times \suinf$, which is the full symmetry of the subsystems. 
This shows the reciprocity of the state of a subsystem and its environment ---any of them can 
be considered as a subsystem or environment. In particular, their entanglement means that they 
have to share at least one common symmetry through which they can be entangled. 
Considering the fact that the finite-rank symmetry $G$ can be different for different 
subsystems, the common symmetry that ensures the global entanglement is necessarily 
$\suinf$. Indeed, considering (\ref{gsuinfcong}), it is possible in $\hvarrho_\infty$ to 
include $\{k_G\}$ parameters into the infinite set of $\suinf$ parameters. Nonetheless, 
the explicit dependence of $\hvarrho_\infty$ on $\{k_G\}$ shows the perspective dependence of 
the environment~\cite{qmrefperspect}. 

\section{Parameter Space of Subsystems}  \label{sec:param}
As the perception of "environment" by different subsystems is not the same, there are an 
infinite number of representations of $\suinf$, one for each subsystem. Moreover, the algebra 
of $ADiff(D_2)$ in (\ref{suinfgendef}) is invariant under scaling of parameters. Thus, the 
area of the diffeo-surface $D_2$ is irrelevant for $\suinfa \cong ADiff(D_2)$. This means 
that (\ref{suinfgendef}) is indeed homomorphic to the algebra 
$\suinfa + \um(1)$~\cite{suninfrep}. 

In the presence of multiple representations of $\suinf$, the homomorphism in (\ref{suinfgdecomp}) 
implies the following relation between the $ADiff$ of their diffeo-surfaces:
\be
\bigcup_{i=1}^n ADiff(D^{(i)}_2) \cong ADiff(D_2), \quad \quad D_2 \equiv 
\bigcup_{i=1}^n  D^{(i)}_2   \label{suinfdiffeo}
\ee
where $D_2$ is, by definition, the diffeo-surface of $\suinf$ in the r.h.s. of 
(\ref{suinfgdecomp}). Although the area of $D_2$ is arbitrary, once diffeo-surfaces are stuck 
together, only the area of their ensemble $D_2$ can be arbitrarily scaled and those of the 
components $D^{(i)}_2$ should be adjusted such that the area of $D_2$ is preserved. Consequently, the 
quantum states of subsystems will depend on an additional continuous parameter $r > 0$ that 
represents the relative area of the compact diffeo-surfaces of their $\suinf$ subsystems, or any 
function of the area, such as its square-roots, as a size indicator. In this interpretation, 
$r = 0$ means a trivial representation of $\suinf$ and, according to the axioms and 
construction of the model, it should be excluded.

Algebraically, (\ref{suinfdiffeo}) is equivalent to breaking of the scaling 
$\mathbb{R} \cong \um(1)$ symmetry of the component subsystems:
\be
\bigotimes^n \biggl (\suinfa + \mathcal{U}(1)\biggr ) ~ 
\autorightarrow{Only total}{area is preserved} \bigotimes^n \suinfa + \mathcal{U}(1) \cong 
\suinfa + \mathcal{U}(1)  \label{suinfadiffeo}
\ee

\begin{figure}
\begin{center}
\includegraphics[width=5cm]{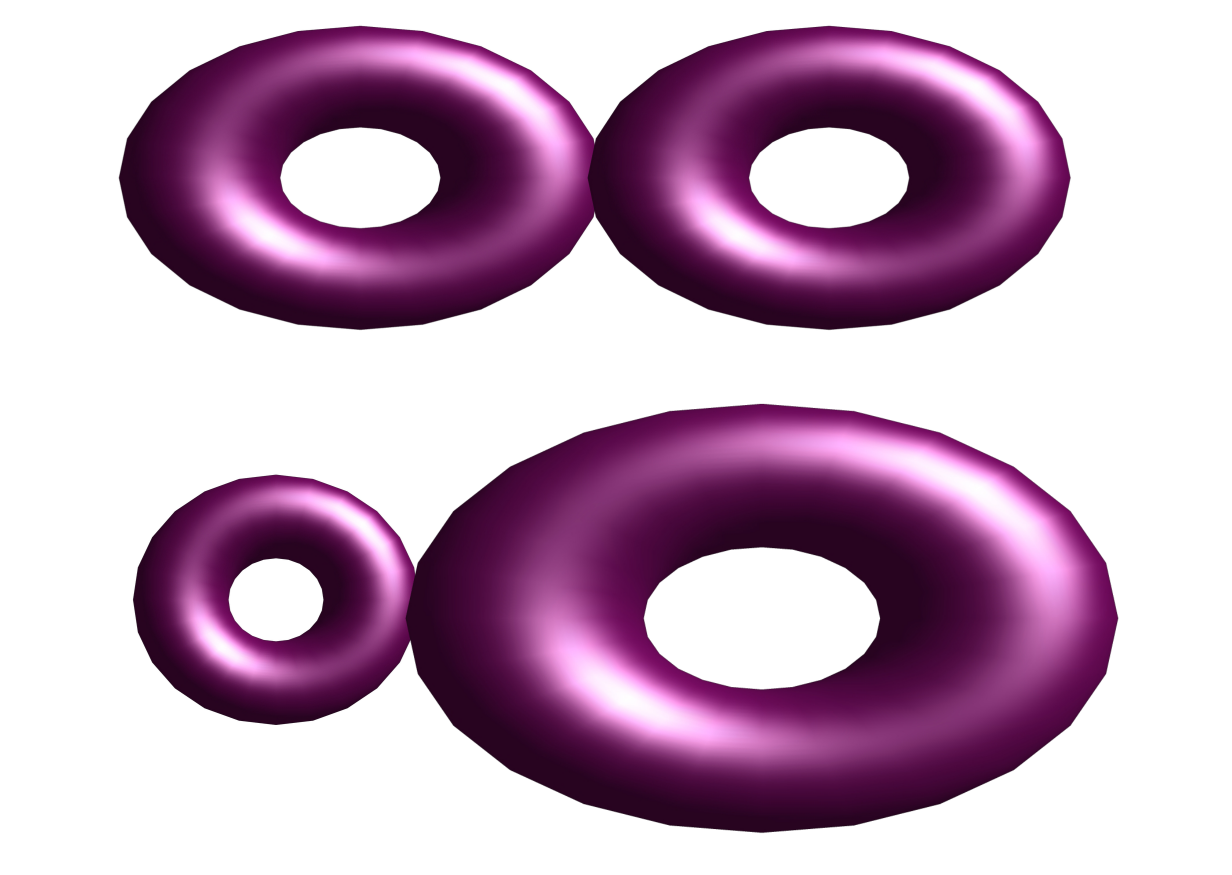}
\end{center}
\caption{Rescaling of the diffeo-surface of left (right) subsystem induces rescaling to the 
right (left) subsystem such that their total area is preserved. \label{fig:diffeoscaling}}
\end{figure}

Diagrams (\ref{singlesuinfdiag}) and (\ref{multisuinfdiag}) summarize the relationship of 
$\suinf$ and $ADiff$ for single and multiple representations. A depiction of the 
diffeo-surfaces of subsystems in Figure \ref{fig:diffeoscaling} shows how the rescaling of 
one affects others, such that the total area is preserved.

\begin{center}
\begin{tabular}{c}
Single subsystem  \\
\begin{tikzcd} 
  \suinf  \dar[swap, "\cong" ] \longrightarrow \suinfa \rar["\text{area irrel.}" ] 
  &  \suinfa + \um (1) \dar["\cong" ] \\
  ADiff(D_2)~ \rar[swap, "\text{area irrel.}"]
  & ~ ADiff(D_2)\times U(1)
\end{tikzcd} \\
\end{tabular}
\be {\label{singlesuinfdiag}} \ee
\end{center}

\begin{center}
\begin{tabular}{c}
Multiple subsystems \\
\begin{tikzcd}[column sep=huge]
\suinf \times \cdots \times \suinf \cong \suinf \longrightarrow \suinfa 
\rar["\text{area irrel.}" ] \dar[swap, "\text{area irrel.}"]  
  & \suinfa + \um (1)  \dar["\text{area irrel.}"] \\
(ADiff(D^{(1)}_2) \times U(1)) \times \cdots \times (ADiff (D^{(n)}_2) \times 
U(1)) \arrow[r, "\text{symm. break}" swap, "\text{area preserv.}"] 
& ADiff(D_2) \times U(1) \\
\end{tikzcd} 
\end{tabular}
\end{center}
\vspace{-0.5cm}
\be \label{multisuinfdiag}  \ee

The breaking of the scaling symmetry does not mean that the value of $r$, or, in other words, 
the areas of diffeo-surfaces of subsystems, are fixed. For each global area and/or its 
rescaling in the r.h.s. of (\ref{multisuinfdiag}), there are infinite number of rescaling 
combinations of the l.h.s. that satisfy the constraint on their total area. Therefore, in 
general, the quantum states of subsystems are in a coherent superposition of $r$ eigen states.

\subsection{Time Parameter and Relative Dynamics}  \label{sec:clocktime} 
In addition to the emergence of an are or size parameter, the division of the Universe into 
subsystems makes it possible to choose one of them as a quantum clock. Then, one of its 
observables can be chosen as a time parameter $t$, and a relative dynamics ---\`a la 
Page and Wootters~\cite{qmtimepage} or equivalent methods~\cite{qmtimedef}--- arises in 
an operational manner as the following: a random application of an operator $\hO$ to the 
state $\hvarrho_s \in {\mathcal B}[{\mathcal H}_s]$ of a subsystem with Hilbert space 
${\mathcal H}_s$ ---in other words, a quantum fluctuation--- changes it to 
$\hO \hvarrho_s \hO^\dagger$. The global entanglement conveys this change to other subsystems, 
in particular to the clock, and its time parameter $t$ 
changes ---"the clock ticks". As the clock state is, by definition, a reference, 
the states of subsystems are tagged by $t$, and variation of their states along with 
the time variation generates a dynamics. 

Of course, the change of the clock's state and, therefore, the time does not need to be 
projective. In addition, other subsystems will have their own change of state, both 
coherently and through reciprocal interactions. Consequently, an arrow of time arises and 
persists eternally because, although inverse processes are, in principle, possible, giving 
the infinite number of subsystems, operators in ${\mathcal B}[{\mathcal H}_s]$, and the 
global entanglement, bringing back the states of all subsystems to their initial 
one ---in other words, inverting the arrow of time--- is extremely improbable. 

\subsection {Properties of Continuous Parameters}  \label{sec:paramsproperty}
We can now complete the list of continuous parameters $(\eta, \zeta, \cdots)$ in 
(\ref{gustate})--(\ref{amptracecond}). With a new ordering, we write them as 
$x \equiv (t, r, \eta, \zeta)$. The last two parameters characterize the representation of 
$\suinf$ by a subsystem and generate its diffeo-surface as a compact 2D subspace $D_2$ of the 
4D parameter space $\Xi \equiv \{x\}$. Parameter $r$ is dimensionful and presents the area 
(or a characteristic length, for example, square-root of area) of the diffeo-surfaces of 
$\suinf$ representations. This is a relative value and is defined with respect to that of 
a reference subsystem. Finally, as we described above, $t$ is a time parameter ---an 
observable of a clock subsystem. 

Although $(t, r, \eta, \zeta)$ parameters have different origins, due to quantum coherence, 
indistinguishability of subsystems having the same symmetries and their representation, and 
arbitrariness of the choice of clock and reference subsystems, they are related to each 
others and the space $\Xi$ cannot be factorized. Thus, diffeo-surfaces of the $\suinf$ 
symmetry of subsystems can be arbitrarily embedded in $\Xi$ and their induced coordinates 
would be functions of all these parameters. Moreover, as described in Section~\ref{sec:symm}, 
in analogy with quantum mechanics on a background spacetime, it is possible to associate 
an operator to each component of $x$ and to their duals. It is 
shown~\cite{houriqmsymmgr,hourisqgr} that these operators satisfy the Heisenberg commutation 
relations. In addition, they can be expanded with respect to the generators $\hL_a$ of 
$\suinf$~\cite{hourisqgr}. However, their expansion is not unique. Consequently, the amount 
of information carried by $\hL_a$ is much larger than what can be expressed by the 
four~observables associated to $x \in \Xi$. 

The observables of subsystems should be invariant or transformable under a reparameterization 
of $\Xi$. On the other hand, a basis transformation of the $\suinf$ sector of the Hilbert 
spaces ${\mathcal H}_s$ is equivalent to a diffeomorphism of the parameter space $\Xi$. 
This is because $\Xi$ can be considered as the collection of 2D 
diffeo-surfaces~\cite{hourisqgr,hourisqgrym}. Inversely, a deformation or reparameterization 
of $\Xi$ can be compensated by a $\suinf$ transformation. As $\suinf \cong ADiff(D_2)$, 
such transformations are equivalent to the deformation of 2D subspaces 
$\Xi$ ---the diffeo-surfaces. Therefore, the geometry of $\Xi$ is irrelevant for 
physical observables. This feature of \sqgr~can be formulated as the following proposition.

\refstepcounter{propos}

\begin{Proposition} \label{propparamcurve}
The curvature of the parameter space $\Xi$ of subsystems can be made trivial by an $\suinf$ 
gauge transformation, under which the Universe and its subsystems are invariant.
\end{Proposition}

In~\cite{hourisqgr}, we used the relationship between Riemann and Ricci curvature tensors, 
Ricci scalar curvature, and the sectional curvature of embedded 2D diffeo-surfaces in $\Xi$ to 
prove this proposition. In~\cite{hourisqgrym}, we confirmed this demonstration by applying 
$\suinf$ transformations on the lowest-order effective Lagrangian of subsystems reviewed 
in Section~\ref{sec:evol}. 

Proposition \ref{propparamcurve} shows that, despite the similarity of $\Xi$ with what we 
perceive as classical spacetime ---especially its dimension--- it cannot be 
identified with the latter, because various astronomical observations have shown the 
influence of the curvature and geometry of the classical spacetime on physical phenomena 
and observables. Nonetheless, in the next section, we find observables that can be identified 
with the classical spacetime.

\section{Classical Geometry as an Effective Path in the Hilbert Space}  \label{sec:geom}
Following the designation of a clock subsystem and a time parameter, the unitary evolution of 
states of subsystems is determined by a Hamiltonian $H_s$ and a Liouvillian operation: 
$d\hvarrho_s/dT = -i/\hbar[\varrho, \hH_s]$. On the other hand, considering Proposition 
\ref{propentang} and global entanglement, the evolution of $\hvarrho_s$ should be more 
realistically formulated by a superoperator 
$\hat{{\mathcal L}} \in {\mathcal B}[{\mathcal B}[{\mathcal H}_s]]$, such that
$d\hvarrho_s/dT = -i/\hbar \hat{{\mathcal L}} (\hvarrho_s)$. The variable $T$ is either the 
outcome of the measurement of the time parameter $t$ of the clock or the expectation value of 
such measurements. In the next section, we explain how the evolution of subsystems are 
formulated, that is, how $\hH_s$ or ${\mathcal L} (\hvarrho_s)$ are determined. Meanwhile, 
we use the Mandelstam--Tamm Quantum Speed Limit (QSL) for the unitary evolution of pure 
states~\cite{qmspeed,qmspeedformul} and analogous relations for the unitary, Markovian, or 
non-Markovian evolution of mixed 
states~\cite{qmspeed,qmspeedgeomopen,qmspeedgeomaction,qmspeedopen,qmspeednonmarkov} to find 
an effective classical spacetime in the framework of \sqgr. We remind readers that QSL relations are 
extensively studied in the literature and the references cited here are only a small sample of 
relevant works.

The QSL inequalities attribute a minimum time to the evolution of a quantum state to another 
completely or partially distinguishable state. The Mandelstam--Tamm QSL 
(MTQSL)~\cite{qmspeed,qmspeedformul} is a consequence of the uncertainty relations between 
non-commuting observables and their unitary evolution according to the Schr\"odinger 
equation. For mixed states and the non-unitary evolution of open systems, geometrical 
properties of the space of density matrices and their relationship with probability 
distributions~\cite{densitymatgeom,densitymatgeom0} provide easier ways to find QSL relations. 
In this approach~\cite{qmspeedgeom,qmspeedgeomopen,qmspeedgeomgen0,qmspeedgeomlocal}, after 
assigning a distance function ${\mathcal D}$ to two states $\hvarrho (T_0)$ and $\hvarrho(T)$ 
and the corresponding metric ${\tt g}_{tt} (\hat{{\mathcal L}}, \hvarrho)$ for the geometry 
of the space of density operators, the QSL can be written as
\be
\Delta T \geqslant \frac{{\mathcal D} (\hvarrho(T_0), \hvarrho(T)}{\llangle \sqrt{{\tt g}_{tt}} 
\rrangle}  \label{ineqgen}
\ee 
where the double bracket means averaging over the measured time interval $\Delta T = T - T_0$ 
along the evolution path in the Hilbert space. However, these QSLs are not all tight, that 
is, the minimum time is not always attainable~\cite{qmspeedgeomlocal,qmspeedrev}. Here, we 
only consider tight QSLs. In any case, the relationship between the geometry of the space 
of density operators and their statistical properties is the evidence that uncertainty 
relations are behind the existence of a speed limit for the evolution of quantum states. 
This is in contrast to the classical physics, in which the speed limit is empirical and an 
axiom of special and general relativity. For example, in the case of pure states, the 
distance ${\mathcal D}$ and the corresponding ${\tt g}_{tt}$ are unique and correspond to the 
Fubini--Study distance and metric, respectively~\cite{qmhilbertgeom}. For mixed states, the 
distance function is not unique and the metric for a given definition of distance is not 
always known~\cite{densitymatgeom,densitymatgeom0,densitymatgeom1}. An exception is the Bures 
distance~\cite{buresmetric,qmspeedgeomopen}, whose corresponding metric is the 
Wigner--Yanase skew information~\cite{wigneryanaseqminfo}. In non-geometric QSL relations, the 
denominator in (\ref{ineqgen}) is usually a non-geometric quantity. For example, for mixed 
states, the relative purity of the state can be used to find an attainable QSL~\cite{qmspeedopen}.

Here, a remark about the physical meaning of time in QSLs is necessary. In the literature, 
the QSL relations, such as (\ref{ineqgen}), are studied in the framework of quantum mechanics 
with a background classical spacetime. By contrast, for \sqgr,~we have to employ them in the 
context of relative time and dynamics, where $T$ should be interpreted as an expectation 
value or conditional outcome of the measurement of the time parameter $t$~\cite{qmtimepage}. 
For this reason, we indicate time as $T$ and not the component $t$ of $x \in \Xi$. Accordingly, 
traces in ${\mathcal D}$ and ${\tt g}_{tt}$ (see~\cite{houriqmsymmgr,hourisqgr} for examples), 
which are calculated at a given time, should take into account the meaning of $T$. For 
instance, it is clear that the tracing operation leading to (\ref{gpurified}) includes all 
components of $x \in \Xi$, including $t$. For using such states in (\ref{ineqgen}), one has 
to project amplitudes on $t = T$ for a projective measurement of time or, more generally, add 
the condition $\tr (\hvarrho \hT) = T$, where $\hT$ is the operator associated to the time 
observable of the clock.

In the framework of \sqgr,~consider an infinitesimal variation of the state of subsystems 
after tracing out the contribution of internal symmetries, that is, 
$\hvarrho_\infty \rightarrow\hvarrho_\infty + d\hvarrho_\infty$. Assume that the clock, its time 
parameter $t$, measured time $T$, and the corresponding evolution superoperator 
$\hat{{\mathcal L}}_\infty$ are such that, in the QSL in (\ref{ineqgen}), equality is achieved:
\be
\llangle \sqrt{{\tt g}_{tt}} \rrangle^2 dT^2 = {\mathcal D}^2 [\hvarrho_\infty, \hvarrho_\infty + 
d\hvarrho_\infty] \equiv ds^2  \label{separation}
\ee
Although (\ref{infdensity}) shows that $\hvarrho_\infty$, $d\hvarrho_\infty$, and, thereby, 
the r.h.s. of (\ref{separation}) are characterized by continuous parameters $x \in \Xi$, it 
also demonstrates that they are independent of $\Xi$'s parameterization. Therefore, the 
introduced parameter $s$ and its variation $ds$ depend only on the state and its variation 
---see the appendices in~\cite{hourisqgr} for an explicit description of $ds$ for the QSL 
examples mentioned above. For this reason (and because of the geometric interpretation of 
QSLs), $s$ is analogous to the affine separation in the classical spacetime. In fact, it is 
indeed the affine separation for the geometry of the space of density 
matrices~\cite{qmspeedgeomopen,qmspeedgeomgen0}.

If we choose another clock with time parameter $t'$,\ and measured time $T'$, the evolution 
superoperator changes to $\hat{{\mathcal L}}'_s$, and, in general, in (\ref{ineqgen}), the 
equality is not attained, because both $\Delta T'$ and the denominator in (\ref{ineqgen}), 
which depends on $\hat{{\mathcal L}}'_s$ (see examples in~\cite{hourisqgr}), change. 
Nonetheless, according to (\ref{separation}), the affine parameter $ds$ only depends on the 
state and its variation. Thus, it remains unchanged, and, in general,
\be
\llangle \sqrt{{\tt g}'_{tt}} \rrangle^2 dT''^2 \geqslant ds^2  \label{separationneq} 
\ee
This inequality can be changed back to equality by adding a term $-d{\mathcal F}^2$ to 
its l.h.s.: 
\be
\llangle \sqrt{{\tt g}'_{tt}} \rrangle^2 dT''^2 - d{\mathcal F}^2 = ds^2  \label{separationdf} 
\ee
To understand the nature of $d{\mathcal F}$, we should remind readers that, for any state $\hvarrho$, 
the affine variation $ds^2$ is a scalar functional of $\hvarrho$ and $d\hvarrho$. Thus, it 
has the general expression $ds^2 = {\mathcal S} [\tr (f_1(\hvarrho)f_2(d\hvarrho)]$, where 
${\mathcal S}$, $f_1$, and $f_2$ are some functions. A tracing operation on the functionals of 
the density operator and its variation is necessary for changing them to C-numbers. Therefore, 
considering the relationship of density matrices with the probability distribution of the outcomes of 
quantum measurements, $ds^2$ presents some sort of statistical averaging. For example, in 
the QSL based on the relative purity~\cite{qmspeedopen} and when Fubini--Study distance is 
used in (\ref{ineqgen}), at the lowest order, $ds^2$ has the following 
form~\cite{hourisqgr,hourisqgrym}: 
\be
ds^2 = (\tr (\hvarrho d\hvarrho))^2  \label{dsave}
\ee
In these cases, $|ds|$ has a clear interpretation as the average variation of state. In the 
context of \sqgr, the state $\hvarrho_\infty$ is characterized by $x \in \Xi$ and the pushback 
of averaging in (\ref{dsave}) leads to an average or effective value $X$ for $x$. We 
call $\Sigma$ the space of these average/expectation values. Notice that both $\hvarrho_\infty$ 
and its purification $|\psi_{G_\infty}\rangle$ in (\ref{gpurified}) are in a superposition of 
$x \in \Xi$ parameters. Therefore, in the framework of quantum mechanics, the average value of 
$x$ is both mathematically and physically meaningful. 

Considering (\ref{separation})--(\ref{dsave}) and the above discussion, we can associate 
a Riemann metric to $\Sigma$ with $s$ as its affine parameter:
\be
ds^2 = g_{\mu\nu} (X) dX^\mu dX^\nu, \mu,~\nu = 0, \cdots, 3  \label{effmetric}
\ee
A reparameterization of $\Xi$, under which observables are invariant, is transferred to 
the space of their expectation values $\Sigma$. Therefore, 4 of 10 components in the metric 
$g_{\mu\nu}(X)$ of $\Sigma$ are arbitrary and we can choose $x$, and thereby $X$, such that 
(\ref{effmetric}) has the following form:
\be
ds^2 = g_{00} (X) dT'^2 - g_{ij}(X) dX^i dX^j, \quad \quad g_{00} (X) = 
\llangle \sqrt{{\tt g}'_{tt}}\rrangle^2 > 0, \quad i,j = 1, 2, 3   \label{effmetric00}
\ee
In this gauge, $X^i,~i= 1, 2, 3$ are related to parameters $(r, \eta, \zeta)$ or, 
equivalently, their Cartesian form $(x^1, x^2, x^3)$. As they are associated to the geometry 
of 2D compact diffeo-surfaces, they and their expectation values $X^i$s should be 
interchangeable. Consequently, \mbox{$g_{ij} (X), i,j = 1, 2, 3$} must have the same sign. 
On the other hand, as, in (\ref{separationdf}), $d{\mathcal F}^2 \geqslant 0$, the 
metric components in (\ref{effmetric00}) must be positive, that is, $g_{ij} (X) > 0$. 
Therefore, the signature of the metric $g_{\mu\nu} (X)$ is negative and $\Sigma$ has a 
Lorentzian (pseudo-Riemannian) geometry. In conclusion, the metric (\ref{effmetric}) has the 
geometrical properties of the classical spacetime. Moreover, Equation~(\ref{dsave}) shows 
that $\Sigma$ and its metric are related to the quantum state of subsystems ---the content 
of the Universe--- and its evolution. For these reasons, we identify the 
(3+1)-dimensional $\Sigma$ with the perceived classical spacetime. In summary, 
\sqgr~explains the origin of both the dimensionality and Lorentzian geometry of the classical 
spacetime. 

If the contribution of internal symmetries in the density matrix is not traced out, one can 
define a metric and an affine parameter that includes parameters of the internal symmetries. 
However, finite-rank Lie groups are usually characterized by discrete parameters. Therefore, 
in contrast to some other QGR proposals, there is no continuous extra-dimension in \sqgr.

\section{Dynamics of Subsystems}  \label{sec:evol}
In the same manner as we did for the whole Universe, we can construct a symmetry-invariant 
Lagrangian on the parameter space $\Xi$ by using symmetry-invariant traces of the products of 
the generators of $\suinf$ and the internal symmetry $G$ of 
subsystems~\cite{houriqmsymmgr,hourisqgr}. The invariance of the coefficients of these traces 
under the reparameterization of $\Xi$ and under \mbox{$G \times \suinf$} symmetry constrains 
their expression, and we find that, at the lowest order in the number of traced generators, 
the effective Lagrangian of subsystems ${\mathcal L}_{U_s}$ must have the form of a 
Yang--Mills theory for both $\suinf$ and $G$ symmetries:
\bea
&& {\mathcal L}_{U_s} = \int d^4x \sqrt{|\upeta|} ~ \biggl [\frac{1}{16\pi L_P^2} 
\tr (F^{\mu\nu} F_{\mu\nu}) + \frac{\lambda}{4} \tr (G^{\mu\nu} G_{\mu\nu}) + 
\frac {1}{2} \sum_s \tr (\sD \hrho_s) \biggr ] \label{yminvarsub} \\
&& F_{\mu\nu} \equiv F_{\mu\nu}^{lm} \hL^{lm} \equiv [D_\mu, D_\nu], \quad D_\mu = 
(\partial_\mu - \Gamma_\mu) \mathbbm{1}_\infty - i \lambda A_\mu, \quad A_\mu, \equiv 
\sum_{lm} A_\mu^{lm} \hL^{lm}, \nonumber \\
&& L_P^2 F_{\mu\nu}^{lm} F^{\mu\nu}_{lm} = L^*_{lm} L^{lm}. \label{yminvardefsuinf} \\
&& G_{\mu\nu} \equiv G_{\mu\nu}^a \hT^a \equiv [D'_\mu, D'_\nu], \quad D'_\mu = 
(\partial_\mu - \Gamma_\mu - i \lambda A_\mu) \mathbbm{1}_G - i \lambda_G B_\mu, \nonumber \\ 
&& B_\mu \equiv \sum_a B_\mu^a \hT^a, \quad L_P^4 G_{\mu\nu}^a G^{\mu\nu}_a = T^*_a T^a. 
\label{yminvardefg}
\eea
The symmetric tensor $\upeta_{\mu\nu}$ is the metric of the parameter space $\Xi$ and should not 
be confused with the metric $g_{\mu\nu}$ of the emergent classical spacetime $\Sigma$. According 
to Proposition \ref{propparamcurve}, $\upeta_{\mu\nu}$ is arbitrary because the geometry of $\Xi$ 
is not a physical observable. The first and second terms in (\ref{yminvarsub}) are the 
Lagrangian density for the $\suinf$ and internal symmetry $G$ gauge fields $A_\mu^{lm}$ and 
$B_\mu^a$, respectively. The $\suinf$ generators $\hL^{lm}$ and range of $(l,m)$ are defined in 
Section~\ref{sec:symm}, and $\Gamma_\mu$ is the geometric connection of the parameter space $\Xi$. 
Operators $T^a$ are the generators of the internal symmetry $G$ of subsystems. Their number is 
determined by the range of index $a$, which must be finite because, as we explain in 
Section~\ref{sec:subsys}, the rank of $G$ is finite. The density matrix $\hrho_s$ is the quantum 
state of a subsystem. The covariant operator $\sD$ is a differential operator and its exact 
definition depends on the representation of symmetries of $\Xi$ by the states of subsystems---see~\cite{hourisqgr} for details and examples. In~\cite{hourisqgrym}, it is shown that geometry 
connection terms in $\sD$ and field equations can be neutralized by an $\suinf$ gauge 
transformation~\cite{hourisqgrym}. Therefore, as mentioned above, the Lagrangian in 
(\ref{yminvarsub}) does not depend on $\upeta_{\mu\nu}$.

A crucial difference between the $\suinf$ sector of ${\mathcal L}_{U_s}$ and $\suinf$ 
Yang--Mills theory on a background spacetime, first studied in~\cite{suninfym}, is that, 
in the latter case the fields depend on two additional continuous parameters, constituting 
so-called ``internal space'' by~\cite{suninfym}. These variables correspond to the 
coordinates of the compact diffeo-surface of $\suinf$ representation. As we discussed 
earlier, in \sqgr,~the parameter space $\Xi$ is indeed the collection of these so-called 
internal spaces or, as we called them, the diffeo-surfaces of subsystems. For this reason, 
they can be identified with two of the four parameters of $\Xi$ or two functions of these 
parameters as the induced coordinates on the diffeo-surface. However, the exact expression 
of these functions is irrelevant, because their variation can be considered as a gauge 
transformation. In addition, Equation~(\ref{yminvardefsuinf}) shows in a simple way why the 
geometry of parameter space $\Xi$ is irrelevant, that is the Proposition \ref{propparamcurve}. 
The geometrical connection $\Gamma_\mu$ of $\Xi$ can be decomposed to spherical harmonic 
functions. Moreover, as discussed in Section~\ref{sec:symm}, the Poisson algebra 
of these functions is homomorphic to $\suinfa$ algebra and can be identified with generators 
$\hL^{lm}$. Therefore, in the covariant derivative $D_\mu$, the geometric connection can be 
included in the $\suinf$ term.

The coefficients of the traces in (\ref{yminvarsub}) can be called ``fields'' because 
they depend on continuous parameters $x \in \Xi$. But they do not need to be quantized 
because, by construction, the effective Lagrangian ${\mathcal L}_{U_s}$ presents the 
lowest-order interactions of a quantum system. They should rather be considered as probability 
distributions. It is evidently trivial to change this quantum mechanical interpretation to a 
QFT one; see the appendices in~\cite{hourisqgrym}. A QFT description would be more useful for 
formulating interactions as a scattering problem, which is useful for testing the model in a
high-energy collider setting. It is important to remind readers that, like all Yang--Mills 
theories, as a QFT, \sqgr~is renormalizable. This is a crucial criterion for any QGR candidate 
model and the main point of failure for many of them. Several other topics, such as parity 
(P), charge conjugate (C), CP symmetries, the possibility of their breaking by adding a 
topological term to the Lagrangian (\ref{yminvarsub}), and the possibility of the existence 
of an ``$\suinf$-axion'', are discussed in~\cite{hourisqgr}.

\section{Classical Limit of Gravity}  \label{sec:limit}
The universal representation of $\suinf$ by all subsystems and their interaction through this 
symmetry according to the Lagrangian ${\mathcal L}_{U_s}$ make $\suinf$ Yang--Mills a good 
candidate for the formulation of quantum gravity, except that, according to observations 
(in particular, the recent detections of gravitational waves), the mediator of classical gravity 
is a spin-2 field. This is in clear contradiction with Yang--Mills gauge theories, in which 
the gauge field is a spin-1 field. In this section, we demonstrate that, if the quantum properties 
of $\suinf$-Yang--Mills are not detectable by the observer ---the case we call the classical 
limit of \sqgr---its effects would be perceived as classical gravity formulated according 
to the Einstein general relativity. A historical parallel of such misperception due to the 
lack of resolution is the sigma model for strong nuclear interaction, in which mesons are the 
mediator particles. Before the discovery of the Quantum Chromo-Dynamics (QCD), the sigma model 
was the favorite because experiments did not have sufficient energy or resolution to detect 
the internal structure of mesons and baryons.

As we discussed in Section~\ref{sec:symm}, if we define the classical limit as 
$\hbar \rightarrow 0$, the \sqgr~becomes trivial and meaningless. This is an expected outcome 
because the model is constructed as being intrinsically quantum, needing $\hbar \neq 0$. 
Moreover, it does not include a classical background spacetime to play the role of a 
``container'' for other entities. For these reasons, we define the classical limit 
as the situation where the observer is not able to detect non-commutative $\suinf$ symmetry 
and associated quantum phenomena. In this case, the observer can only perceive (measure) the 
expectation values $X \in \Sigma$ of the parameters $x$ defined in (\ref{effmetric}). 
Consequently, the terms in the action in (\ref{yminvarsub}) are perceived as functions of $X$, 
and integration should be performed over the effective space $\Sigma$ using its Lorentzian 
metric $g_{\mu\nu}$. Moreover, in the absence of information about the $\suinf$ gauge field, 
the first term in 
${\mathcal L}_{U_s}$ is perceived as a scalar function on the manifold $\Sigma$. According to 
Theorem 4.35 in~\cite{curvaturfunc}, every scalar function defined on a (pseudo-)Riemannian 
manifold with dimension $D \geq 3$ is the scalar curvature for a (pseudo-)Riemannian metric. 
Therefore, when the structure of the $\suinf$ gauge term in (\ref{yminvarsub}) is not 
discernible, it can be considered as being proportional to the scalar curvature of a metric, 
specifically that of $g_{\mu\nu}$. Therefore, in the classical limit, as defined above, the 
Lagrangian ${\mathcal L}_{U_s}$ is perceived as
\be
{\mathcal L}_{U_s} ~ \autorightarrow{Classical}{limit} ~ {\mathcal L}_{cl.gr} = 
\int_{\Sigma} d^4X \sqrt{|g|} \biggl [\kappa R^{(4)} + \frac{1}{4} \tr (G^{\mu\nu} G_{\mu\nu}) + 
\frac {1}{2} \sum_s \tr (\sD \hrho_s) \biggr ] \label{classicgr}
\ee
where the dimensionful constant $\kappa \propto M_P^2$ is necessary to make the pure gravity 
term dimensionless. This Lagrangian has the form of $G$-symmetry Yang--Mills QFT with matter in 
a classical curved spacetime. It is also clear that other possible scalars obtained from 
the curvature tensor of the same metric, such as $R^n,~ n > 1$, or $R^{\mu\nu} R_{\mu\nu}$ should be 
related to higher quantum orders ---not included in the lowest-order actions in (\ref{yminvarsub}) 
and (\ref{classicgr}). They are smaller than the Einstein--Hilbert term by a factor of at least 
${\mathcal O}(1) \hbar^2/M_P^2$; hence, currently undetectable. We also remind readers that 
the definition of $g_{\mu\nu}$ in (\ref{effmetric}) and the relationship of the affine parameter 
with the quantum state of subsystems in (\ref{dsave}) do not allow one to determine $g_{\mu\nu}$. 
Therefore, there is no conflict or duplicity in the calculation of effective metric $g_{\mu\nu}$. 

Because we interpret the $\suinf$ sector of (\ref{yminvarsub}) as gravity, it is important to 
check whether it has the same number of degrees of freedom (d.o.f.) as the classical Einstein 
gravity. It is easy to see that they are indeed the same. In a space of dimension $D$, the 
antisymmetric 2-form $F_{\mu\nu}\equiv F_{\mu\nu}^{lm} \hL^{lm}$ has $D(D-1)/2$ components. This 
field is reparameterization-invariant, meaning that it is independent of the connection 
$\Gamma_\mu$. In classical gravity $g_{\mu\nu}$ (or, equivalently, the Ricci tensor $R_{\mu\nu}$) 
is a symmetric tensor and has $D$ degrees of redundancy due to reparameterization. This leaves 
$D(D +1)/2 - D = D(D-1)/2$ independent d.o.f. Therefore, as tensor quantities on a parameter 
space or on a classical spacetime, they have the same number of degrees of freedom and there 
is no inconsistency in this respect between \sqgr~and its classical limit as Einstein gravity. 
Thus, a fundamental spin-1 mediator for quantum gravity is not in contradiction with the 
observed classical spin-2 graviton. Indeed, the equality of the d.o.f. is one of the reasons 
for the gauge-gravity duality conjecture used in the construction of some QGR 
models~\cite{spaceemerge,qgrgaugedual,stringgauge1}. Also notice that modified gravity models 
such as $F(T),~ F(Q),~ F(T,Q)$, which are recently 
proposed as a modified teleparallel formulation of the classical gravity ---see, 
e.g.,~\cite{grteleparal} for a review--- depend on additional d.o.f. for torsion and 
nonmetricity. Consequently, they break the required equality of the number of d.o.f. in the 
quantum and classical limit of \sqgr. Moreover, the Besse theorem used for obtaining the 
classical limit action (\ref{classicgr}) is proven for the (pseudo-)Riemannian manifold 
without torsion. Therefore, in the framework of \sqgr,~additional terms such as 
$F(R),~ F(T),~ F(Q),~ F(T,Q)$ in the lowest order classical limit effective action of gravity 
cannot be fundamental. Nonetheless, as we explain below, they may arise as special 
configurations of $\suinf$ fields. 

The classical gravity Lagrangian ${\mathcal L}_{cl.gr}$ does not include a cosmological 
constant, and there is no trivial candidate in \sqgr~that add such a term to 
${\mathcal L}_{U_s}$ or ${\mathcal L}_{cl.gr}$. However, considering the inconsistencies between 
the measured cosmological parameters from the Cosmic Microwave Background 
(CMB)~\cite{cmbplanckparam} and those estimated from late Universe probes such as supernovae, 
micro-lensing, and Baryon Acoustic Oscillation (BAO) ---known as ``Hubble and $S_8$ 
tensions'' in the literature~\cite{hubbletensionrev}--- dark energy may be dynamical 
rather than a cosmological constant. Indeed, results of the BAO measurement by the DESI 
survey~\cite{lssbaodesi1y} seem to be more consistent with an evolving dark energy density 
rather than a cosmological constant. Nonetheless, in~\cite{hourisqgr}, a few processes are 
suggested that may generate an effective cosmological constant in \sqgr. In particular, 
in~\cite{hourisqgr}, it is suggested that topologically nontrivial $\suinf$ field 
configurations such as instantons may behave as a dynamical dark energy or an effective 
modified gravity. In fact, it is shown that the main difference between modified classical 
gravity $F(R),~ F(T),~ F(Q),~ F(T,Q)$ models, which are also dark energy candidates, is the 
boundary configuration of the metric $g_{\mu\nu}$~\cite{modgrboundary}. Non-trivial 
configurations of the $\suinf$ gauge field $A^{\mu}_{lm}$ can be the fundamental quantum origin 
for what may classically be interpreted as torsion or nonmetricity and, thereby, modified 
classical gravity. In particular, it may be possible to embed the $\mathbb{R}^4$ Abelian 
gauge field of teleparallel formulation in the $\suinf$ gauge field. A thorough study of 
these possibilities is left to future works. Finally, many other dark energy models studied 
in the literature are also relevant in the \sqgr~framework.

\section{Concluding Remarks}  \label{sec:conclu}
As a candidate model for quantum gravity and cosmology, \sqgr~differs both in its construction 
and predictions from other QGR proposals. It is constructed axiomatically as a quantum system 
and its postulates include neither an interaction similar to gravity nor a spacetime or fields 
that play similar roles, as is the case in string theory. All these concepts are emergent. 
The most distinctive prediction of \sqgr~is a spin-1 mediator boson at a quantum level for the 
interaction that is classically perceived as gravity with a spin-2 mediator. On the other hand, the 
origin and properties of this spin-1 mediator diverge from those in gauge-gravity duality 
models~\cite{spaceemerge,qgrgaugesep0,qgrgaugesep1,gaugestringcorr,stringgauge0,spaceemerge,qgrgaugedual,stringgauge1}, 
or the gauging of translation symmetry in the frame-based classical gravity 
models, such as teleparallelism and its extensions. 
The $\suinf$ gauge symmetry is not directly 
related to Lorentz invariance and the diffeomorphism of the perceived classical spacetime. 

Considering the differences between \sqgr~and what we know about gravity and our classical perception 
of spacetime, the question arises as to whether it is a viable model for quantum cosmology and gravity. 
Our answer is that it is a plausible model because it is self-consistent and, despite its 
abstract axioms, it approaches Einstein gravity 
and some of its extensions
, in the sense discussed in Section~\ref{sec:limit}. 
Given the fact that, at present, there is no observed evidence of QGR and its properties, any 
intrinsically consistent model that predicts Einstein gravity in the classical limit is a 
potentially viable candidate. As we explained in the Introduction, some of the popular QGR 
candidates, which seem closer and more inspired by what we know about gravity, do not go as far 
as the newcomer and abstract \sqgr~in the demonstration of their consistency with the Einstein 
gravity and testable predictions.

Future works should investigate predictions of \sqgr~for processes and phenomena in which 
quantum gravity is considered to be important, such as the quantum structure of black holes and 
the puzzle of apparent information loss in these objects, inflation, particle physics beyond 
the standard model, and laboratory tests of quantum gravity. 
Another important 
topic for future works is an in-depth study of \sqgr-specific dark energy models and their 
relation with phenomenological models, such as modified gravity proposals.

\vspace{6pt}
\funding{This research received no external funding.}

\dataavailability{Data are contained within the article.} 

\conflictsofinterest{The author declares no conflict of interest. }

\begin{adjustwidth}{-\extralength}{0cm}

\reftitle{References}


\PublishersNote{}
\end{adjustwidth}


\begin{thebibliography}{999}
\bibitem {admgr} \Name{R.}{Arnowitt}, \Name{S.}{Deser}, \Name{C.}{Misner}, ~\emph{Dynamical Structure and Definition of Energy in General Relativity}, \href{https://journals.aps.org/pr/abstract/10.1103/PhysRev.116.1322}{\Journal{\PRV}{116}{1959}{1322}}, [\href{https://arxiv.org/abs/gr-qc/0405109}{arXiv:gr-qc/0405109}].
\bibitem {ashtekarvar} \Name{A.}{Ashtekar}, ~\emph{New Variables for Classical and Quantum Gravity}, \href{https://journals.aps.org/prl/abstract/10.1103/PhysRevLett.57.2244}{\Journal{\PRL}{57}{1986}{2244}}.
\bibitem {spinfoamlimit} \Name{F.}{Markopoulou}, ~\emph{Coarse graining in spin foam models}, \href{https://iopscience.iop.org/article/10.1088/0264-9381/20/5/301}{\Journal{\CQG}{20}{2003}{777}}, \href{https://arxiv.org/abs/gr-qc/0203036}{[arXiv:gr-qc/0203036]}.
\bibitem {spinfoamlimit0} \Name{D.E.}{Neville}, ~\emph{On the Classical Limit of Spin Network Gravity: Two Conjectures}, \href{https://arxiv.org/abs/0807.1043}{[arXiv:0807.1043]} (2008).
\bibitem {spinfoamlimit1} \Name{M.}{Han}, ~\emph{Einstein Equation from Covariant Loop Quantum Gravity in Semiclassical Continuum Limit}, \href{https://journals.aps.org/prd/abstract/10.1103/PhysRevD.96.024047}{\Journal{\PRD}{96}{2017}{024047}}, \href{https://arxiv.org/abs/1705.09030}{[arXiv:1705.09030]}.
\bibitem {grestgrb090510a} Fermi GBM/LAT Collaborations:\Name{A.A.}{Abdo}, \Name{M.}{Ackermann}, \Name{M.}{Ajello}, \Name{K.}{Asano}, \Name{W.B.}{Atwood}, \Name{M.}{Axelsson}, \Name{L.}{Baldini}, \Name{J.}{Ballet}, \Name{G.}{Barbiellini}, \Name{M.G.}{Baring}, \etal, ~\emph{A limit on the variation of the speed of light arising from quantum gravity effects}, \href{https://www.nature.com/articles/nature08574}{\Journal {\NAT}{462}{2009}{331}}, [\href{http://arxiv.org/abs/0908.1832}{arXiv:0908.1832}].  
\bibitem {lorentzviolconstr} The LHAASO Collaboration, ~\emph{Stringent Tests of Lorentz Invariance Violation from LHAASO Observations of GRB 221009A}, \href{https://journals.aps.org/prl/abstract/10.1103/PhysRevLett.133.071501}{\Journal{\PRL}{133}{2024}{071501}}, \href{https://arxiv.org/abs/2402.06009}{[arXiv:2402.06009]}.  
\bibitem {spacetimeemerge} \Name{N.}{Seiberg}, ~\emph{Emergent Spacetime}, in proceedings of \href{https://www.worldscientific.com/doi/abs/10.1142/9789812706768_0005}{The Quantum Structure of Space and Time, (2007), 163}, \href{https://arxiv.org/abs/hep-th/0601234}{[arXiv:hep-th/0601234]}. 
\bibitem {grclassic} \Name{J.}{Oppenheim}, ~\emph{A post-quantum theory of classical gravity?}, (2018) [\href{http://arxiv.org/abs/1811.03116}{arXiv:1811.03116}]. 
\bibitem {qmentanglemb} \Name{H.}{Li}, \Name{J.-P.}{Dou}, \Name{X.-L.}{Pang}, \Name{C.-N.}{Zhang}, \Name{Z.-Q.}{Yan}, \Name{T.-H.}{Yang}, \Name{J.}{Gao}, \Name{J.-M.}{Li}, \Name{X.-M.}{Jin}, ~\emph{Multipartite entanglement of billions of motional atoms heralded by single photon}, \href{https://www.nature.com/articles/s41534-021-00476-1}{\Journal{\em Nature Quantum Info}{7}{2021}{146}}.
\bibitem {qmentanglemm} \Name{R.}{Riedinger}, \Name{A.}{Wallucks}, \Name{I.}{Marinkovic}, \Name{C.}{Löschnauer}, \Name{M.}{Aspelmeyer}, \Name{S.}{Hong}, \Name{S.}{Gröblacher}, ~\emph{Remote quantum entanglement between two micromechanical oscillators}, \href{https://arxiv.org/abs/1710.11147}{\Journal {\NAT}{556}{2018}{473}}, \href{https://arxiv.org/abs/1710.11147}{[arXiv:1710.11147]}. 
\bibitem {qgrlocalqm1} \Name{S.B.}{Giddings}, ~\emph{Quantum-first gravity}, \href{https://link.springer.com/article/10.1007/s10701-019-00239-1}{\Journal {\FOP}{49}{2019}{177}}, [\href{http://arxiv.org/abs/1803.04973}{arXiv:1803.04973}].
\bibitem {qgrhistory} \Name{J.B.}{Hartle} ~ Generalizing Quantum Mechanics for Quantum Spacetime. In \emph{The Quantum Structure of Space and Time};  Gross, D., Henneaux, M.,  Sevrin, A., Eds.; World Scientific: Singapore, 2007, \href{https://arxiv.org/abs/gr-qc/0602013}{[arXiv:gr-qc/0602013]}.
\bibitem {qgrlocalqm0} \Name{W.}{Donnelly} ; \Name{S.B.}{Giddings} ~ How is quantum information localized in gravity? \Journal{\PRD}{96}{2017}{086013}, \href{https://arxiv.org/abs/1706.03104}{[arXiv:1706.03104]}.
\bibitem {qgrlocalqm2} \Name{W.}{Donnelly} ; \Name{S.B.}{Giddings} ~ Gravitational splitting at first order: Quantum information localization in gravity. \Journal{\PRD}{98}{2018}{086006}, \href{https://arxiv.org/abs/1805.11095}{[arXiv:1805.11095]}.
\bibitem {qgrentangle1} \Name{M.}{Van Raamsdonk}, ~\emph{Lectures on Gravity and Entanglement}, in Proceeding of \href{https://www.worldscientific.com/worldscibooks/10.1142/10270?srsltid=AfmBOooRXm7b5cDBbrGacTe9tZxxcVC_qHslyxMxhJfpphmMU_7VJRAV#t=aboutBook}{\emph{New Frontiers in Fields and Strings}, TASI 2015}, pp. 297-351, World Scientific (2017), \href{https://arxiv.org/abs/1609.00026}{[arXiv:1609.00026]}, (2016).
\bibitem {qgrentangle2} \Name{C.}{Cao} ; \Name{S.M.}{Carroll} ; \Name{S.}{Michalakis} ~ Space from Hilbert Space: Recovering Geometry from Bulk Entanglement. \Journal{\PRD}{95}{2017}{024031}, \href{http://arxiv.org/abs/1606.08444}{[arXiv:1606.08444]}.
\bibitem {qgrentangle3} \Name{C.}{Cao} ; \Name{S.M.}{Carroll} ~ Bulk Entanglement Gravity without a Boundary: Towards Finding Einstein's Equation in Hilbert Space. \Journal{\PRD}{97}{2018}{086003}, \href{https://arxiv.org/abs/1712.02803}{[arXiv:1712.02803]}.
\bibitem {houriqmsymmgr} \Name{H.}{Ziaeepour}, ~\emph{Making a Quantum Universe: Symmetry and Gravity}, \href{https://www.mdpi.com/2218-1997/6/11/194}{\Journal{\MDU}{6(11)}{2020}{194}}, [\href{https://arxiv.org/abs/2009.03428}{arXiv:2009.03428}].
\bibitem {hourisqgr} \Name{H.}{Ziaeepour}, ~\emph{\boldmath $\suinf$-QGR: Emergence of Gravity in an Infinitely Divisible Quantum Universe}, [\href{https://arxiv.org/abs/2301.02813}{arXiv:2301.02813}].
\bibitem {hourisqgrym} \Name{H.}{Ziaeepour}, ~\emph{Quantum state of fields in $\mathbf{\suinf}$ Quantum Gravity}, [\href{https://arxiv.org/abs/2402.18237}{arXiv:2402.18237}], (submitted).
\bibitem {hourisqgrcomp} \Name{H.}{Ziaeepour}, ~\emph{Comparing Quantum Gravity Models: Loop Quantum Gravity, Entanglement and AdS/CFT versus $SU(\infty)$-QGR}, \href{http://dx.doi.org/10.3390/sym14010058}\href{https://www.mdpi.com/2073-8994/14/1/58}{\Journal{\SYM}{14}{2022}{58}}, [\href{https://arxiv.org/abs/2109.05757}{arXiv:2109.05757}].
\bibitem {houriqmsymm} \Name{H.}{Ziaeepour}, ~\emph{Symmetry as a foundational concept in Quantum Mechanics}, \href{https://iopscience.iop.org/article/10.1088/1742-6596/626/1/012074/meta}{\Journal{\JPC}{626}{2015}{012074}}, [\href{http://arxiv.org/abs/1502.05339}{arXiv:1502.05339}].
\bibitem {suninfrep0}\Name{Y.}{Zunger}, ~\emph{Why Matrix theory works for oddly shaped membranes}, \href{https://journals.aps.org/prd/abstract/10.1103/PhysRevD.64.086003}{\Journal{\PRD}{64}{2001}{086003}}, [\href{http://arxiv.org/abs/hep-th/0106030}{arXiv:hep-th/0106030}].
\bibitem {suninfhoppthesis} \Name{J.}{Hoppe}, ~\emph{Quantum Theory of a Massless Relativistic Surface and a Two-dimensional Bound State Problem}, \href{https://dspace.mit.edu/handle/1721.1/15717}{Ph.D. Thesis, MIT}, Cambridge, MA, USA, (1982).
\bibitem {suninftorus} \Name{J.}{Hoppe}, ~\emph{Diffeomorphism Groups, Quantization, and $SU(\infty)$}, \href{https://www.worldscientific.com/doi/abs/10.1142/S0217751X89002235}{\Journal {\IMA}{4}{1989}{5235}}.
\bibitem{suninfsurfaceanomal} \Name{T.A.}{Arakelyan}, \Name{G.K.}{Savvidy}, ~\emph{Cocycles of area-preserving diffeomorphisms and anomalies in the theory of relativistic surfaces}, \href{https://www.sciencedirect.com/science/article/abs/pii/0370269388913755}{\Journal{\PLB}{214}{1988}{350}}.
\bibitem{suinftriang} \Name{D.B.}{Pairlie}, \Name{P.}{Fletcher}, \Name{C.K.}{Zachos}, ~\emph{Trigonometric Structure Constants for New Infinite-Dimensional Algebras}, \href{https://www.sciencedirect.com/science/article/abs/pii/0370269389914184}{\Journal{\PLB}{218}{1989}{203}}.
\bibitem{suinftriang0} \Name{D.B.}{Pairlie}, \Name{C.K.}{Zachos}, ~\emph{Infinite-dimensional algebras, sine brackets, and $\suinf$}, \href{https://www.sciencedirect.com/science/article/abs/pii/0370269389910575}{\Journal{\PLB}{224}{1989}{101}}.
\bibitem{suninfrep} \Name{J.}{Hoppe}, ~\Name{P.}{Schaller}, ~\emph{Infinitely Many Versions of $SU(\infty)$}, \href{https://www.sciencedirect.com/science/article/abs/pii/037026939091197J}{\Journal {\PLB}{237}{1990}{407}}.
\bibitem {qmmathbook} \Name{B.C.}{Hall}, ~\emph{Quantum Theory for Mathematicians}, Springer, (2013). 
\bibitem {qgrnoncommut} \Name{A.}{Connes}, ~\emph{Gravity coupled with matter and foundation of non-commutative geometry}, \href{https://link.springer.com/article/10.1007/BF02506388}{\Journal{\CMP}{182}{1996}{155}}, [\href{https://arxiv.org/abs/hep-th/9603053}{arXiv:hep-th/9603053}].
\bibitem {prodstate} \Name{A.}{Peres}, ~\emph{Separability Criterion for Density Matrices}, \href{https://journals.aps.org/prl/abstract/10.1103/PhysRevLett.77.1413}{\Journal{\PRL}{77}{1996}{1413}} \href{https://arxiv.org/abs/quant-ph/9604005}{[quant-ph/9604005]}. 
\bibitem {sysdiv} \Name{P.}{Zanardi}, ~\Name{D.}{Lidar}, ~\Name{S.}{Lloyd}, ~\emph{Quantum tensor product structures are observable-induced}, \href{https://journals.aps.org/prl/abstract/10.1103/PhysRevLett.92.060402}{\Journal{\PRL}{92}{2004}{060402}}, [\href{http://arxiv.org/abs/quant-ph/0308043}{arXiv:quant-ph/0308043}].
\bibitem {qminfocohere} \Name{T.}{Baumgratz}, \Name{M.}{Cramer}, \Name{M.B.}{Plenio}, ~\emph{Quantifying Coherence}, \href{https://journals.aps.org/prl/abstract/10.1103/PhysRevLett.113.140401}{\Journal {\PRL}{113}{2014}{140401}}, [\href{http://arxiv.org/abs/1311.0275}{arXiv:1311.0275}]. 
\bibitem {spherharmonbound} \Name{P.}{Garrett}, ~\emph{Harmonic analysis on spheres}, Lecture Notes, (2014).
\bibitem {qminfobook} \Name{M.A.}{Nielsen}, \Name{I.L.}{Chuang}, ~\emph{Quantum Computation and Quantum Information}, Cambridge University Press, (2001).
\bibitem {qmpurifcond} \Name{M.}{Kleinmann}, \Name{H.}{Kampermann}, \Name{T.}{Meyer}, \Name{D.}{Bruss}, \emph{Physical Purification of Quantum States}, \href{https://journals.aps.org/pra/abstract/10.1103/PhysRevA.73.062309}{\Journal{\PRA}{73}{2006}{062309}}, \href{https://arxiv.org/abs/quant-ph/0509100}{[quant-ph/0509100]}.
\bibitem {qmpurif} \Name{E.}{Schr\"odinger}, ~\emph{Probability relations between separated systems}, Proceedings of the Cambridge Philosophical Society. 32 (1936) 446.
\bibitem {qmpurif0} \Name{N.}{Hadjisavvas}, ~\emph{Properties of mixtures on non-orthogonal states}, \href{https://link.springer.com/article/10.1007/BF00401481}{\Journal{Lett.Math.Phys.}{5}{1981}{327}}.
\bibitem {qmpurif1} \Name{L.P.}{Hughston}, \Name{R.}{Jozsa}, \Name{W.K.}{Wootters}, (November 1993). \emph{A complete classification of quantum ensembles having a given density matrix}, \href{https://www.sciencedirect.com/science/article/pii/0375960193908809}{\Journal{\PLA}{183}{1993}{14}}. 
\bibitem {qmrefperspect} \Name{S.}{Ali Ahmad}, \Name{T.D.}{Galley}, \Name{P.A.}{H\"ohn}, \Name{M.P.E.}{Lock}, \Name{A.R.H.}{Smith}, ~\emph{Quantum Relativity of Subsystems}, \href{https://doi.org/10.1103/PhysRevLett.128.170401}{\Journal{\PRL}{128}{2022}{170401}}, \href{https://arxiv.org/abs/2103.01232}{[arXiv:2103.01232]}.
\bibitem {qmtimepage} \Name{D.N.}{Page}, ~\Name{W.K.}{Wootters}, ~\emph{Evolution without evolution: Dynamics described by stationary observables}, \href{https://journals.aps.org/prd/abstract/10.1103/PhysRevD.27.2885}{\Journal{\PRD}{27}{1983}{2885}}.
\bibitem {qmtimedef} \Name{P.A.}{Hoehn}, ~\Name{A.R.H.}{Smith}, ~\Name{M.P.E.}{Lock}, ~\emph{The Trinity of Relational Quantum Dynamics}, \href{https://journals.aps.org/prd/abstract/10.1103/PhysRevD.104.066001}{\Journal{\PRD}{104}{2021}{066001}}, [\href{http://arxiv.org/abs/1912.00033}{arXiv:1912.00033}].
\bibitem {qmspeed} \Name{L.}{Mandelstam}, ~\Name{I.}{Tamm}, ~\emph{The Uncertainty Relation Between Energy and Time in Non-relativistic Quantum Mechanics}, \href{https://link.springer.com/chapter/10.1007/978-3-642-74626-0_8}{\Journal{\JPU}{9}{1945}{249}}.
\bibitem {qmspeedformul} \Name {K.}{Bhattacharyya}, -\emph{Quantum decay and Mandelstam-Tamm energy inequality}, \href{https://iopscience.iop.org/article/10.1088/0305-4470/16/13/021}{\Journal{\JPA: {\em Math. Gen}}{16}{1983}{2993}}.
\bibitem {qmspeedgeomopen} \Name{A}{Uhlmann}, -\emph{An energy dispersion estimate}, \href{https://www.sciencedirect.com/science/article/abs/pii/037596019290555Z}{\Journal{\PLA}{161}{1992}{329}}.  
\bibitem {qmspeedgeomaction} \Name{K.}{Funo}, \Name{N.}{Shiraishi}, \Name{K.}{Saito}, ~\emph{Speed limit for open quantum systems}, \href{https://iopscience.iop.org/article/10.1088/1367-2630/aaf9f5}{\Journal{\NJP}{21}{2019}{013006}}, [\href{https://arxiv.org/abs/1810.03011}{arXiv:1810.03011}].
\bibitem {qmspeedopen} \Name{A.}{del Campo}, \Name{I.L.}{Egusquiza}, \Name{M.B.}{Plenio}, \Name{S.F.}{Huelga}, ~\emph{Quantum speed limits in open system dynamics}, \href{https://journals.aps.org/prl/abstract/10.1103/PhysRevLett.110.050403}{\Journal{\PRL}{110}{2013}{050403}}, [\href{https://arxiv.org/abs/1209.1737}{arXiv:1209.1737}]. 
\bibitem {qmspeednonmarkov} \Name{S.}{Deffner}, \Name{E.}{Lutz}, ~\emph{Quantum speed limit for non-Markovian dynamics}, \href{https://journals.aps.org/pre/abstract/10.1103/PhysRevE.77.021128}{\Journal{\PRL}{111}{2013}{010402}}, \href{https://arxiv.org/abs/1302.5069}{[arXiv:1302.5069]} 
\bibitem {densitymatgeom} \Name{E.A.}{Morozova}, \Name{N.N.}{Chentsov}, ~\emph{Markov invariant geometry on manifolds of states}, \href{https://link.springer.com/article/10.1007/BF01095975}{\Journal{\em Journal of Soviet Math.}{56}{1991}{2648}}.
\bibitem {densitymatgeom0} \Name{D.}{Petz}, \Name{H.}{Hasegawa}, ~\emph{Metric of $\alpha$-Entropies of Density Matrices}, \href{https://link.springer.com/article/10.1007/BF00398324}{\Journal{\em Lett. Math. Phys.}{38}{1996}{221}}. 
\bibitem {qmspeedgeom} \Name {J.}{Anandan}, \Name{Y.}{Aharonov}, ~\emph{Geometry of quantum evolution}, \href{https://journals.aps.org/prl/abstract/10.1103/PhysRevLett.65.1697}{\Journal{\PRL}{65}{1990}{1697}}.
\bibitem {qmspeedgeomgen0}\Name{D.}{Paiva Pires}, \Name{M.}{Cianciaruso}, \Name{L.C.}{Céleri}, \Name{G.}{Adesso}, \Name{D.O.}{Soares-Pinto}, ~\emph{Generalized Geometric Quantum Speed Limits}, \href{https://journals.aps.org/prx/abstract/10.1103/PhysRevX.6.021031}{\Journal {\PRX}{6}{2016}{021031}}, [\href{http://arxiv.org/abs/1507.05848}{arXiv:1507.05848}].
\bibitem {qmspeedgeomlocal} \Name{E.}{O'Connor}, \Name{G.}{Guarnieri}, \Name{S.}{Campbell}, ~\emph{Action quantum speed limits}, \href{https://journals.aps.org/pra/abstract/10.1103/PhysRevA.103.022210}{\Journal {\PRA}{103}{2021}{022210}}, [\href{http://arxiv.org/abs/2011.05232}{arXiv:2011.05232}]
\bibitem {qmspeedrev} \Name{S.}{Deffner}, \Name{S.}{Campbell}, ~\emph{Quantum speed limits: from Heisenberg's uncertainty principle to optimal quantum control}, \href{https://iopscience.iop.org/article/10.1088/1751-8121/aa86c6}{\Journal{\JPA}{50}{2017}{453001}}, [\href{https://arxiv.org/abs/1705.08023}{arXiv:1705.08023}].
\bibitem {qmhilbertgeom} \Name{G.W.}{Gibbons}, ~\emph{Typical states and density matrices}, \href{https://www.sciencedirect.com/science/article/abs/pii/0393044092900464}{\Journal{\em J. Geometry \& Phys.}{8}{1992}{147}}.
\bibitem {densitymatgeom1} \Name{D.}{Petz}, ~\emph{Monoton Matrices on Matrix spaces}, \href{}{\Journal{\em Linear Algebra Appl.}{244}{1996}{81}}.
\bibitem {buresmetric} \Name{D.}{Bures}, ~\emph{An Extension of Kakutant's Theorem on Infinite Product Measures to the Tensor Product of Semifinite $w^*$-Algebras}, \href{https://www.jstor.org/stable/1995012?casa_token=BCgMgaFHRY4AAAAA\%3AbFS9omYbO_tgJ3KEu8x8WBX3W-ej4KPdwhZh7ysWB6YPhVqwgpyH326sZxcStnwp0bdugN00Rs1Y-OzpJbQcOv_MzzzD2G3Fby_bP9FqgoeNM4aADF0}{\Journal{\em Trans.Am.Math.Soc}{135}{1969}{199}}. 
\bibitem {wigneryanaseqminfo} \Name{E.P.}{Wigner}, \Name{M.M.}{Yanase}, ~\emph{Information Content of Distributions}, \href{https://www.pnas.org/doi/abs/10.1073/pnas.49.6.910}{\Journal{\em Proc. Nation. Acad. Sci. USA}{49}{1963}{910}}.
\bibitem {suninfym} \Name{E.G.}{Floratos}, ~\Name{J.}{Iliopoulos}, ~\Name{G.}{Tiktopoulos}, ~\emph{A note on $SU(\infty)$ classical Yang-Mills theories}, \href{https://www.sciencedirect.com/science/article/abs/pii/0370269389908678}{\Journal{\PLB}{217}{1989}{285}}. 
\bibitem {curvaturfunc} \Name{A.L.}{Besse}, ~\emph{Einstein manifolds}, in \emph{Results in Mathematics and Related Areas (3)}, Springer-Verlag, Berlin, (1987).

\bibitem {spaceemerge} \Name{F.}{Wilczek} ~ \emph{Riemann--Einstein Structure from Volume and Gauge Symmetry}. \href{https://journals.aps.org/prl/abstract/10.1103/PhysRevLett.80.4851}{\Journal {\PRL}{80}{1998}{4851}}, \href{https://arxiv.org/abs/hep-th/9801184}{[arXiv:hep-th/9801184]}.
\bibitem {qgrgaugedual} \Name{T.}{Banks} ; \Name{W.}{Fischler} ; \Name{S.H.}{Shenker} ; \Name{L.}{Susskind}  ~ M Theory As A Matrix Model: A Conjecture. \Journal{\PRD}{55}{1997}{5112}, \href{https://arxiv.org/abs/hep-th/9610043}{[arXiv:hep-th/9610043]}.
\bibitem {stringgauge1} \Name{O.}{Aharony}, \Name{S.S.}{Gubser}, \Name{J.}{Maldacena}, \Name{H.}{Ooguri}, \Name{Y.}{Oz}, ~\emph{Large N Field Theories, String Theory and Gravity}, \href{https://www.sciencedirect.com/science/article/abs/pii/S0370157399000836}{\Journal{\PRE}{323}{2000}{183}}, [\href{https://arxiv.org/abs/hep-th/9905111}{arXiv:hep-th/9905111}].
\bibitem{grteleparal} \Name{J.W.}{Maluf}, \emph{The teleparallel equivalent of general relativity}, \href{https://onlinelibrary.wiley.com/doi/10.1002/andp.201200272}{\Journal{\APN}{525}{2013}{339}}, [\href{https://arxiv.org/abs/1303.3897}{arXiv:1303.3897}].
\bibitem {cmbplanckparam} \Name{Planck Collaboration}{}, ~\emph{Planck 2018 results. VI. Cosmological parameters}, \href{https://www.aanda.org/articles/aa/full_html/2020/09/aa33910-18/aa33910-18.html}{\Journal{\AA}{641}{2020}{A6}}, [\href{https://arxiv.org/abs/1807.06209}{arXiv:1807.06209}].
\bibitem {hubbletensionrev} \Name{E.}{Abdalla}, \Name{G.F.}{Abellán}, \Name{A.}{Aboubrahim}, \Name{A.}{Agnello}, \Name{O.}{Akarsu}, \Name{Y.}{Akrami}, \Name{G.}{Alestas}, \Name{D.}{Aloni}, \etal, ~\emph{Cosmology Intertwined: A Review of the Particle Physics, Astrophysics, and Cosmology Associated with the Cosmological Tensions and Anomalies}, \href{https://www.sciencedirect.com/science/article/pii/S2214404822000179}{\Journal {\JHA}{34}{2022}{49}}, [\href{http://arxiv.org/abs/2203.06142}{arXiv:2203.06142}].
\bibitem{lssbaodesi1y} \Name{DESI Collaboration}{}, ~\emph{DESI 2024 VI: Cosmological Constraints from the Measurements of Baryon Acoustic Oscillations}, [\href{https://arxiv.org/abs/2404.03002}{arXiv:2404.03002}].
\bibitem {modgrboundary} \Name{C.G.}{Boehmer}, \Name{E.}{Jensko}, ~\emph{Modified gravity: a unified approach}, \href{https://journals.aps.org/prd/abstract/10.1103/PhysRevD.104.024010}{\Journal{\PRD}{104}{2021}{024010}} , [\href{https://arxiv.org/abs/2103.15906}{arXiv:2103.15906}]. 
\bibitem {qgrgaugesep0} \Name{A.}{Torres-Gomez}, \Name{K.}{Krasnov}, ~\emph{Gravity-Yang-Mills-Higgs unification by enlarging the gauge group}, \href{https://journals.aps.org/prd/abstract/10.1103/PhysRevD.81.085003}{\Journal{\PRD}{81}{2010}{085003}}, [\href{http://arxiv.org/abs/0911.3793}{arXiv:0911.3793}]. 
\bibitem {qgrgaugesep1} \Name{J.W.}{Barrett}, \Name{S.}{Kerr}, ~\emph{Gauge gravity and discrete quantum models}, (2013), [\href{http://arxiv.org/abs/1309.1660}{arXiv:1309.1660}]. 
\bibitem {gaugestringcorr} \Name {S.S.}{Gubser}, \Name {I.R.}{Klebanov}, \Name {A.M.}{Polyakov}, ~\emph{Gauge Theory Correlators from Non-Critical String Theory}, \href{https://www.sciencedirect.com/science/article/abs/pii/S0370269398003773}{\Journal{\PLB}{428}{1998}{105}}, [\href{https://arxiv.org/abs/hep-th/9802109}{arXiv:hep-th/9802109}].
\bibitem {stringgauge0} \Name{E.}{Witten}, ~\emph{Anti De Sitter Space And Holography}, \href{https://www.intlpress.com/site/pub/pages/journals/items/atmp/content/vols/0002/0002/a002/}{\Journal{\ATM}{2}{1998}{253}}, [\href{https://arxiv.org/abs/hep-th/9802150}{arXiv:hep-th/9802150}].

\end{thebibliography}
\end{document}